\documentclass[%
 reprint,
superscriptaddress,
 amsmath,amssymb,
 aps,
]{revtex4-2}

\usepackage{graphicx}
\usepackage{dcolumn}
\usepackage{bm}

\usepackage{comment}
\usepackage{braket}
\usepackage{bbold}
\usepackage{comment}
\usepackage{float}
\usepackage{hyperref}
\usepackage{academicons}
\usepackage{xcolor}

\begin{document}

\preprint{APS/123-QED}

\title{Demonstration of long-range correlations via susceptibility measurements in a one-dimensional superconducting Josephson spin chain}

\author{D. M. Tennant}
\thanks{Present Address: Lawrence Livermore National Laboratory, Livermore, California 94550, USA}
\email{tennant3@llnl.gov}
\affiliation{Institute for Quantum Computing, University of Waterloo, Waterloo, ON, Canada N2L 3G1}
\affiliation{Department of Physics and Astronomy, University of Waterloo, Waterloo, ON, Canada N2L 3G1}


\author{X. Dai}
\affiliation{Institute for Quantum Computing, University of Waterloo, Waterloo, ON, Canada N2L 3G1}
\affiliation{Department of Physics and Astronomy, University of Waterloo, Waterloo, ON, Canada N2L 3G1}

\author{A. J. Martinez}
\affiliation{Institute for Quantum Computing, University of Waterloo, Waterloo, ON, Canada N2L 3G1}
\affiliation{Department of Physics and Astronomy, University of Waterloo, Waterloo, ON, Canada N2L 3G1}

\author{R. Trappen}
\affiliation{Institute for Quantum Computing, University of Waterloo, Waterloo, ON, Canada N2L 3G1}
\affiliation{Department of Physics and Astronomy, University of Waterloo, Waterloo, ON, Canada N2L 3G1}

\author{D. Melanson}
\affiliation{Institute for Quantum Computing, University of Waterloo, Waterloo, ON, Canada N2L 3G1}
\affiliation{Department of Physics and Astronomy, University of Waterloo, Waterloo, ON, Canada N2L 3G1}

\author{M. A. Yurtalan}
\affiliation{Institute for Quantum Computing, University of Waterloo, Waterloo, ON, Canada N2L 3G1}
\affiliation{Department of Physics and Astronomy, University of Waterloo, Waterloo, ON, Canada N2L 3G1}

\author{Y. Tang}
\affiliation{Institute for Quantum Computing, University of Waterloo, Waterloo, ON, Canada N2L 3G1}
\affiliation{Department of Physics and Astronomy, University of Waterloo, Waterloo, ON, Canada N2L 3G1}

\author{S. Bedkihal}
\affiliation{Institute for Quantum Computing, University of Waterloo, Waterloo, ON, Canada N2L 3G1}
\affiliation{Department of Physics and Astronomy, University of Waterloo, Waterloo, ON, Canada N2L 3G1}

\author{R. Yang}
\affiliation{Institute for Quantum Computing, University of Waterloo, Waterloo, ON, Canada N2L 3G1}
\affiliation{Department of Physics and Astronomy, University of Waterloo, Waterloo, ON, Canada N2L 3G1}

\author{S. Novikov}
\affiliation{Northrop Grumman Corporation, Linthicum, Maryland 21090, USA}

\author{J. A. Grover}
\affiliation{Northrop Grumman Corporation, Linthicum, Maryland 21090, USA}

\author{S. M. Disseler}
\affiliation{Northrop Grumman Corporation, Linthicum, Maryland 21090, USA}

\author{J. I. Basham}
\affiliation{Northrop Grumman Corporation, Linthicum, Maryland 21090, USA}

\author{R. Das}
\affiliation{MIT Lincoln Laboratory, Lexington, Massachusetts 02421, USA}

\author{D. K. Kim}
\affiliation{MIT Lincoln Laboratory, Lexington, Massachusetts 02421, USA}

\author{A. J. Melville}
\affiliation{MIT Lincoln Laboratory, Lexington, Massachusetts 02421, USA}

\author{B. M. Niedzielski}
\affiliation{MIT Lincoln Laboratory, Lexington, Massachusetts 02421, USA}

\author{S. J. Weber}
\affiliation{MIT Lincoln Laboratory, Lexington, Massachusetts 02421, USA}

\author{J. L. Yoder}
\affiliation{MIT Lincoln Laboratory, Lexington, Massachusetts 02421, USA}

\author{A. J. Kerman}
\affiliation{MIT Lincoln Laboratory, Lexington, Massachusetts 02421, USA}

\author{E. Mozgunov}
\affiliation{Center for Quantum Information Science \& Technology, University of Southern California, Los Angeles, California 90089, USA}

\author{D. A. Lidar}
\affiliation{Center for Quantum Information Science \& Technology, University of Southern California, Los Angeles, California 90089, USA}
\affiliation{Departments of Electrical \& Computer Engineering, Chemistry, and Physics, University of Southern California, Los Angeles, California 90089, USA}

\author{A. Lupascu}
\email{adrian.lupascu@uwaterloo.ca}
\affiliation{Institute for Quantum Computing, University of Waterloo, Waterloo, ON, Canada N2L 3G1}
\affiliation{Department of Physics and Astronomy, University of Waterloo, Waterloo, ON, Canada N2L 3G1}
\affiliation{Waterloo Institute for Nanotechnology, University of Waterloo, Waterloo, ON, Canada N2L 3G1}



\date{\today}

\begin{abstract}
Spin chains have long been considered an effective medium for long-range interactions, entanglement generation, and quantum state transfer.  In this work, we explore the properties of a spin chain implemented with superconducting flux circuits, designed to act as a connectivity medium between two superconducting qubits.  The susceptibility of the chain is probed and shown to support long-range, cross chain correlations.  In addition, interactions between the two end qubits, mediated by the coupler chain, are demonstrated.  This work has direct applicability in near term quantum annealing processors as a means of generating long-range, coherent coupling between qubits.

\end{abstract}

\maketitle
 

\section{\label{sec:level1}Introduction}
Superconducting quantum information platforms have reached a level of maturity where tens of individual qubits, comprising a computational device, can provide proof of principle demonstrations of quantum simulations, quantum algorithms, and basic error correction functionality \cite{kjaergaard20}.  As these devices, and the tasks they seek to address, scale in size and complexity, so does the need for realizing qubit networks with increased dimensionality and expanded connectivity.  These two desired features of future quantum processors prompt the development of long-range, qubit coherence preserving interactions \cite{katzgraber14,bravyi21}.  Quantum spin chains have been proposed as an effective medium for qubit interactions with these desired properties \cite{venuti06,friesen07,venuti07,ferreira08,oh10,oh11}.  In this article, we explore the possibility of long-range interactions supported by quantum spin chains for superconducting qubits \cite{weber18,kerman18, kerman20}.  This architecture has direct application in recently proposed quantum annealing platforms based on superconducting capacitively shunted flux qubits \cite{weber17,novikov18}, rf-SQUIDs \cite{harris10}, fluxmon qubits \cite{quintana17}, and fluxonium qubits \cite{nguyen19}. 

Quantum annealing is emerging as a promising paradigm for near term quantum computing \cite{apolloni88,das08,hauke20,albash18}.  An initial Hamiltonian, whose ground state is straightforward to prepare, is transformed continuously to the problem Hamiltonian.  The prepared state of the problem Hamiltonian is located in the vicinity of the true ground state and represents a useful solution of the optimization problem.  In the limit of weak coupling to the environment, adiabatic quantum computing has been shown to be immune to dephasing in the energy basis, making it a particularly attractive candidate for near term, noisy quantum computing platforms \cite{albash15}.  Commercial quantum annealers, based on superconducting Josephson flux qubits \cite{johnson11, lanting14, harris18}, have recently become available to the larger community and are beginning to make their mark as a valuable research tool, see e.g., Refs. \citenum{ikeda19,inoue21,li18}.

There are strong motivations for improving upon the performance of quantum annealing processors \cite{katzgraber18}, in particular with respect to how their constituent qubits interact with one another.  Increasing both the graph dimensionality of qubit networks \cite{katzgraber14,heim15}, and improving connectivity \cite{venturelli15}, the degree to which the qubits are coupled to one another, would greatly reduce physical hardware overhead by increasing the types and sizes of optimization problems that can be natively embedded.  Existing quantum annealing processors based on superconducting qubits possess either nearest neighbor \cite{novikov21} or a combination of inter- and intra- unit cell interactions \cite{boothby20} between qubits.  Commercial annealers, possessing this combination of inter- and intra- unit cell interactions, currently rely on minor embedding \cite{choi08,choi11}, a procedure of extending logical qubits over multiple physical qubits to implement problems that require higher dimensionality or connectivity than the processor's hardware natively allows.  

\begin{figure*}[t]
\centering
\includegraphics[width = \linewidth]{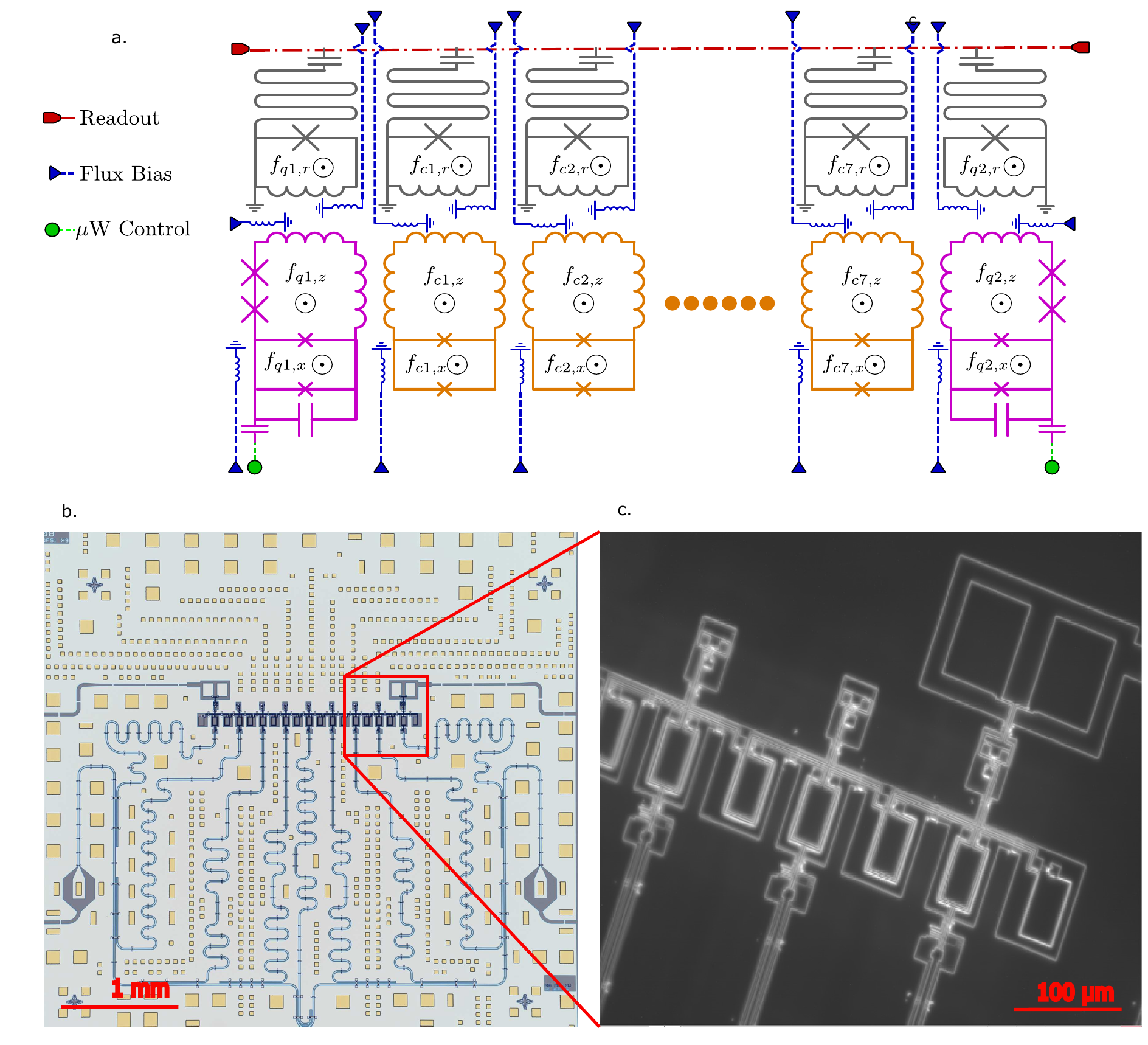}
\caption{\textbf{Description of the Device.}  \textbf{a)}  A schematic showing the full device and circuit geometry: two end qubits, shown in magenta, and seven coupler units, shown in orange, each equipped with individual readout resonators.  Also indicated are the flux control lines for the qubits, couplers, and readout resonators as well as the microwave control lines for the qubits and the microwave feed-through line for state readout.  \textbf{b) \& c)}  Optical images of the device chip and an expanded view of one end of the coupler chain.  The end qubit, capacitively coupled qubit microwave control line, the two adjacent coupler units, and respective readout structures are shown expanded in Panel \textbf{(c)}.  The readout rf-SQUIDs, connecting to both the qubit and coupler units, terminate at the end of their meandering resonators.  The flux control lines are located on the opposing interposer tier.} \end{figure*}

As each connection made to a qubit introduces additional noise and decoherence channels, expanding qubit connectivity in quantum annealing processors must be balanced against the need to maintain the qubits' coherence properties.  Developing quantum annealing processors that support improved qubit coherence times would allow greater functionality in computation.  Higher precision flux control, afforded by improved coherence, is required by many computational problems of interest \cite{lucas14,ohkuwa18, passarelli19}.  In general, just how much of a computational advantage greater qubit coherence provides in quantum annealing processes is itself an open scientific question \cite{denchev16,albash2018}.  Furthermore, more coherent quantum annealers will enable diabatic annealing protocols that require a greater degree of qubit coherence throughout the annealing process \cite{somma12,munoz19,crosson21}.  

These two, often competing, improvements - creating qubit networks with higher dimensionality and expanded connectivity and maintaining qubit coherence - call for further development of long-range qubit interactions.  One proposed scheme that accomplishes this dual need is utilizing spin chains as the qubit interaction medium \cite{venuti06,venuti07,ferreira08}.  Gapped spin chains, in the context of semiconducting quantum dots \cite{friesen07,oh10,oh11}, have been proposed to support long-range, Ruderman-Kittel-Kasuya-Yoshida (RKKY) type qubit interactions \cite{ruderman54}.  Recent progress in this direction includes a demonstration of adiabatic quantum state transfer along a linear array of four electron spin qubits \cite{qiao21}.  In addition to the possibility of supporting coherent coupling between two distant qubits, the spin chain architecture lends itself to higher connectivity schemes.  Multiple qubits can be simultaneously interacting with a single 1-D chain \cite{friesen07}.  Additionally, paramagnetic trees, formed by spin chains forking into multiple paths, offer another possible scheme for higher qubit connectivity \cite{weber18,kerman18,kerman20}.   

This work demonstrates the viability of this coupling scheme in the context of superconducting Josephson qubit hardware.  In the following we discuss long-range coupling mediated by quantum spin chains in a hardware independent fashion.  This is a more natural language to describe long-range coupling as a consequence of the system's underlying quantum phase transition \cite{osborne02,vidal03,ai08,qpt_note}.  Following this discussion, we demonstrate a realization of the quantum spin model with superconducting circuits.  To accomplish this, we design a system of two qubits, coupled together through a chain of seven spin units.  The spin chain, shown in Fig. 1, is realized by a one-dimensional array of seven tunable rf-SQUIDs \cite{brink05,zakosarenko07,harris07,allman10,allman14,weber17} inductively coupled to their nearest neighbor through the SQUIDs' main loops.  Each end coupler is inductively coupled to a tunable, capacitively shunted, superconducting flux qubit \cite{orlando99,van00,paauw09,yan16}.  Finally, to illustrate the viability of mediating long-range, coherent qubit interactions with our device, we characterize the non-local susceptibility of the coupler chain, demonstrate long-range qubit-qubit interactions, and identify the parameter region where both long-range correlations exist and the detrimental effects of low frequency flux noise are negligible. 

The Hamiltonian for the quantum spin chain is the one-dimensional Ising model.  Incorporating the two end qubits, it can be written as
\begin{equation}
    H = H_{\rm{q}} + H_{\rm{c}} + H_{\rm{int}},
\end{equation}
with 
\begin{equation}
H_{\rm{q}} = \sum\limits_{i=1}^2(\frac{\epsilon_{\rm{q}_i}}{2}\sigma^z_{\rm{q}_i} + \frac{\Delta_{\rm{q}_i}}{2}\sigma^x_{\rm{q}_i}),
\end{equation}
\begin{equation}
H_{\rm{c}} = \sum\limits_{i=1}^7(\frac{\epsilon_{\rm{c}_i}}{2}\sigma^z_{\rm{c}_i} + \frac{\Delta_{\rm{c}_i}}{2}\sigma^x_{\rm{c}_i}) + \sum\limits_{i=1}^6 J_{\rm{c}_i \rm{c}_ {i+1}}\sigma^z_{\rm{c}_i}\sigma^z_{\rm{c}_{i+1}},
\end{equation}
%
and
\begin{equation}
H_{\rm{int}} = J_{\rm{q}_1 \rm{c}_1}\sigma^z_{\rm{q}_1}\sigma^z_{\rm{c}_1} + J_{\rm{q}_2 \rm{c}_7}\sigma^z_{\rm{c}_7}\sigma^z_{\rm{q}_2}.
\end{equation}
In the previous equations, $\Delta_{\rm{q}_i} /2$ ($\Delta_{\rm{c}_i} /2$) and $\epsilon_{\rm{q}_i} /2$ ($\epsilon_{\rm{c}_i} /2$) are the transverse and longitudinal components of the qubits' (couplers') spin while $J_{\rm{c}_i \rm{c}_{i+1}}$ and $J_{\rm{q}_i \rm{c}_j}$ represent the coupling strength between adjacent coupler units and between qubits and their nearest coupler unit.  For the remainder of the article, we will assume the coupler units are operated homogeneously, that is $\epsilon_{\rm{c}_i}\,=\,\epsilon_{\rm{c}}$, $\Delta_{\rm{c}_i}\,=\,\Delta_{\rm{c}}$ and $J_{\rm{c}_i \rm{c}_{i+1}}\,=\, J_{\rm{cc}}$.  

Virtual excitations of the coupler chain can be integrated over to derive an expression for the coupler-chain-mediated effective qubit-qubit interaction strength, $J^{\rm{eff}}_{\rm{q}_1 \rm{q}_2}$.  By considering the qubit-adjacent coupler unit interaction, $J_{\rm{q}_i \rm{c}_j}$, to be a weak perturbation to the coupler Hamiltonian, the interaction energy can be calculated to second order as the shift of the ground state energy of the coupler Hamiltonian.  As the operating temperature of the device will be much less than the coupler chain excitation energy, it is reasonable to assume that the coupler chain remains in its ground state and the cross chain interactions are supported by virtual excitations \cite{venuti07,oh11,sm}.  This is reminiscent of the RKKY interaction whose long-range interaction between magnetic impurities is mediated by virtual excitations of conduction electrons above the Fermi surface \cite{ruderman54}.  

\begin{figure}[t]
\centering
\includegraphics[width = \linewidth]{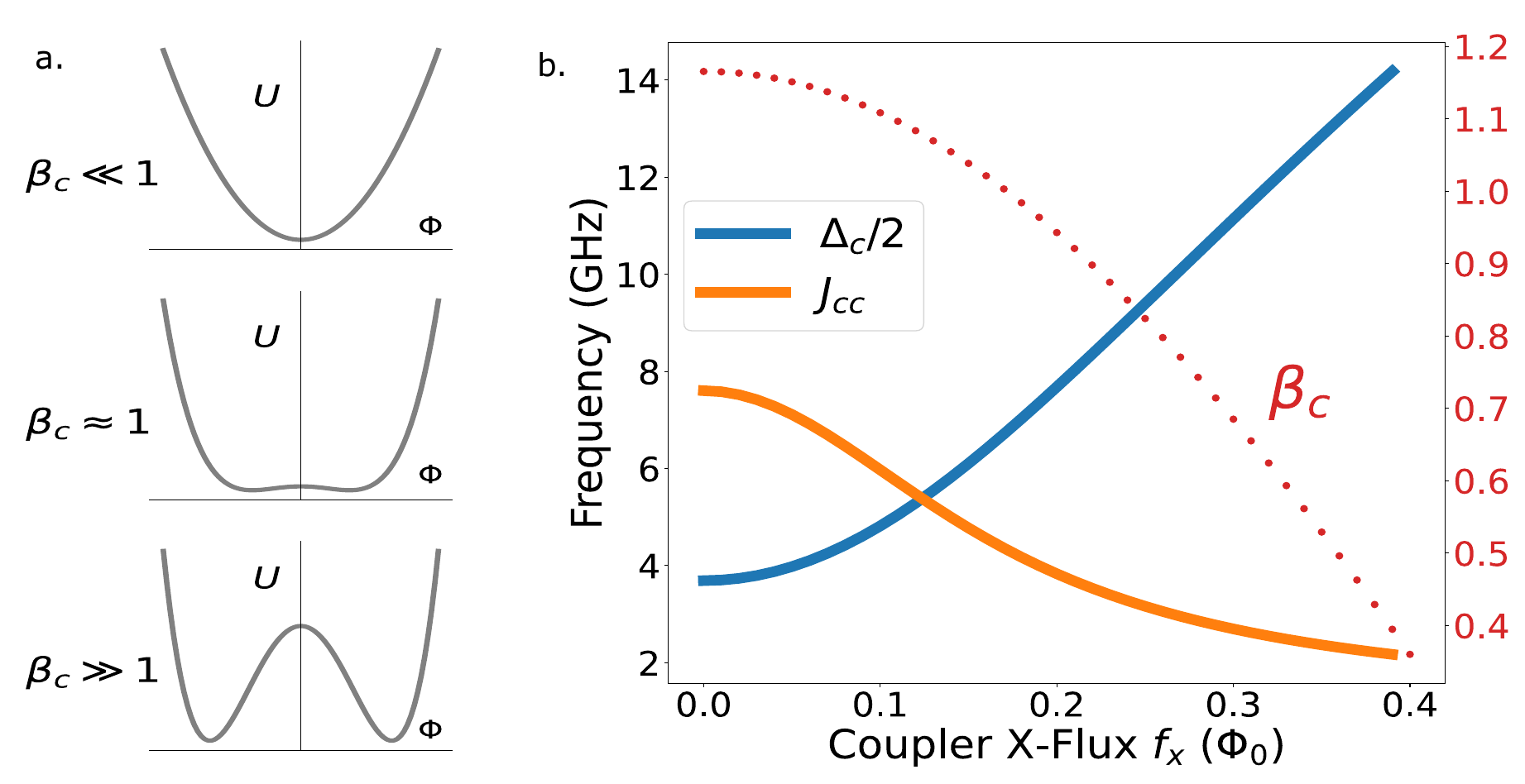}
\caption{\textbf{Single Coupler Behavior.} \textbf{a)}  The ratio of the Josephson inductance to the geometrical inductance, $\beta_c$, dictates the shape of the potential energy of the tunable rf-SQUID coupler circuit.  When the geometrical inductance dominates, $\beta_{\rm{c}} \ll 1$, the potential energy landscape is approximately harmonic.  When the Josephson inductance dominates, $\beta_{\rm{c}} \gg 1$, the energy landscape becomes double-welled with each minimum representing oppositely circulating current states.  Due to the large energy barrier between the states, moderate changes in $f_z$ do not change the current state of the circuit.  The coupling is optimized when the geometrical and Josephson inductances are approximately equal.  This results in a wide, shallow energy minimum where even slight changes in $f_z$ can induce strong fluctuations between the oppositely circulating current states of the coupler circuit.  \textbf{b)}  Both the single coupler transverse field, $\Delta_{\rm{c}}/2$, and the intercoupler interaction energy, $J_{\rm{c}_i \rm{c}_{i+1}} = M_{\rm{c}_i \rm{c}_{i+1}}I^z_{\rm{c}_i} I^z_{\rm{c}_{i+1}}$, are displayed as a function of the coupler $f_x$ when $f_z = \Phi_0/2$.  These parameters are calculated in single coupler simulations and then transcribed into spin model parameters.  The equality of these two terms appearing in the transverse field Ising model for $f_x \simeq 0.14\, \Phi_0$ signals the location of the quantum critical point, in the vicinity of which we expect long-range correlations to emerge.  In addition, the dependence of $\beta_{\rm{c}}$ is displayed as a function of the coupler's $f_x$ for $f_z = \Phi_0/2$.  By design, the optimum coupling point, $\beta_{\rm{c}} \approx 1$, coincides with the coupler $f_x$ value where we expect critical behavior in the coupler chain.}
\end{figure}

By taking these above stated approximations into account it is possible to derive an expression for the chain mediated effective coupling strength between the end qubits \cite{sm}.  The effective Hamiltonian is
\begin{equation}
\begin{split}
    H&_{\rm{q}_1, \rm{q}_2}^{\rm{eff}} = H_{\rm{q}} + J_{\rm{q}_1 \rm{q}_2}^{\rm{eff}}\sigma^z_{\rm{q}_1}\sigma^z_{\rm{q}_2}, \\
    J^{\rm{eff}}_{\rm{q}_1 \rm{q}_2} \approx \,& \frac{J_{\rm{q}_1 \rm{c}_1} J_{\rm{q}_2 \rm{c}_7}}{\Omega_{\rm{c}}}(\bra {0_{\rm{c}}} \sigma^z_{\rm{c}_1} \ket{0_{\rm{c}}} \bra {0_{\rm{c}}} \sigma^z_{\rm{c}_7} \ket{0_{\rm{c}}} \\
    & - \bra {0_{\rm{c}}} \sigma^z_{\rm{c}_1} \sigma^z_{\rm{c}_7} \ket{0_{\rm{c}}}),
\end{split}    
\end{equation}
where $\Omega_{\rm{c}}$ is the energy gap between the coupler chain ground and first excited state and $\ket{0_{\rm{c}}}$ represents the collective ground state of the unperturbed seven unit coupler chain.  Note that an exact expression for the effective coupling between qubits contains the integrals of frequency dependent connected coupler correlation functions. The ground state connected correlation function in Eq. (5) is an approximation assuming a large excitation gap, $\Omega_{\rm{c}}$, and that the coupler chain excitation frequencies are sufficiently degenerate \cite{ferreira08,oh10}. This approximation is strictly valid in the case where the transverse field on each coupler unit is much larger than the exchange interaction between coupler units.  Writing this expression in terms of the zero temperature, bulk susceptibility of the response function in the Lehmann representation, $\widetilde{\chi}_{\rm{c}_1 \rm{c}_7}$, \cite{ferreira08} the effective interaction can be expressed as
\begin{equation}
    J^{\rm{eff}}_{\rm{q}_1 \rm{q}_2} = \widetilde{\chi}_{\rm{c}_1 \rm{c}_7}\,J_{\rm{q}_1 \rm{c}_1}\,J_{\rm{q}_2 \rm{c}_7}. 
\end{equation}

The main objective of this work is to measure the quantity $\widetilde{\chi}_{\rm{c}_i \rm{c}_j}$ as a function of $J_{\rm{cc}}/(\Delta_{\rm{c}} / 2)$, the ratio of the inter-coupler longitudinal coupling strength, proportional to $\sigma^z_{\rm{c}_i}\sigma^z_{\rm{c}_{i+1}}$, to the individual coupler unit transverse field strength, oriented along $\sigma^x_{\rm{c}_i}$, for the homogeneously tuned chain.  Long-range coherent coupling becomes possible when the spin chain is tuned to the vicinity of its quantum critical point \cite{osborne02,vidal03}.  In the case presented here, this occurs when the strength of the transverse fields of the coupler spins and inter-unit longitudinal coupling energies between the nearest-neighbor coupler spins become comparable.  The coupler chain susceptibility is determined by measuring the response of the longitudinal fields of the coupler units along the chain when a small longitudinal field, $\delta\epsilon$, is applied to the end coupler unit.  It is shown that the response truly becomes long-range, that is entirely cross chain, for $J_{\rm{cc}}/(\Delta_{\rm{c}} / 2) \gtrsim 1$, where the system approaches and enters its ordered phase.  

\begin{figure*}
\centering
\includegraphics[width = \linewidth]{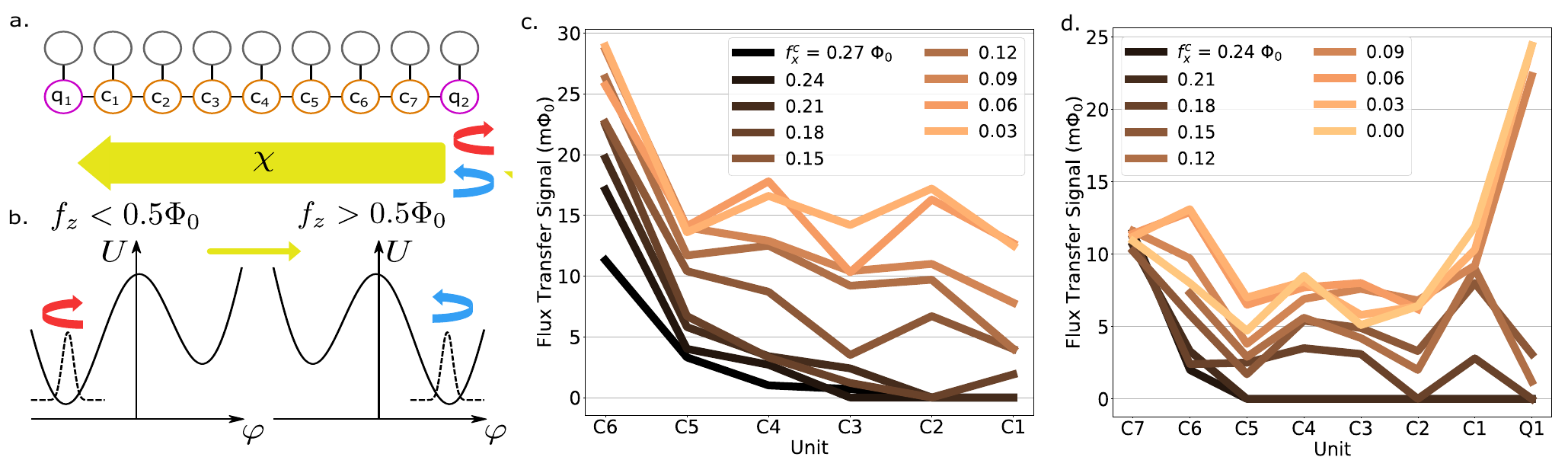}
\caption{\textbf{The Flux Propagation Experiment.}  \textbf{a)}  A schematic depicting the flux propagation experiment.  Either Coupler 7 [results in (\textbf{c})], or Qubit 2 [results in (\textbf{d})], act as the source unit generating the flux signal.  The flux response is then measured in the different units along the chain.  \textbf{b)}  The flux signal is generated by sweeping the $f_z$ of the source unit across its one-half magnetic flux bias point causing that unit's circulating current to change direction.  \textbf{c) \& d)}  The magnitude of the experimentally measured flux signal is displayed for different homogeneously tuned coupler $f_x$ values for the coupler-only (\textbf{c}) and full device (\textbf{d}).  The displayed signal is the difference in effective $f_z = \Phi_0/2$ point when the source unit is placed at $f_z = \Phi_0/2 \pm 20\, \rm{m}\Phi_0$.  As the strong coupling between the tunable resonator and the qubit (coupler) makes it challenging to model the resonator response, we therefore resort to image processing techniques to determine the effective symmetry point of the qubit (coupler) unit.  This sets the uncertainty of the extracted flux signal to be on the order of the pixel size of the scan, 2.4 m$\Phi_0$.  In the case of both the coupler only and qubit to qubit measurements, the cross chain signal becomes non-zero in the vicinity of the coupler chain's predicted critical region.}
\end{figure*}

The one-dimensional transverse field Ising spin model can be realized by multilevel superconducting Josephson circuits.  With the assumption of negligible state occupation of higher levels, the two lowest energy levels of the circuit define the qubit subspace where the transverse and longitudinal components of the unit's spin can be determined as a function of the applied magnetic flux.  The individual qubit and coupler circuits utilize inductive couplings for implementing the inter-unit interactions $J_{\rm{q}_1 \rm{c}_1}$, $J_{\rm{c}_7 \rm{q}_2}$, and $J_{\rm{c}_i \rm{c}_{i+1}}$.  Coupling of this type for single unit coupler circuits has been demonstrated in flux qubits \cite{brink05,zakosarenko07,harris07,weber17}, phase qubits \cite{allman10,allman14}, and fluxmons \cite{quintana17}.  The design choice of independent coupler circuits, as opposed to direct coupling between qubits, is particularly appealing for use in annealing processors where it is necessary to independently control the qubit properties and coupling strengths.  This single unit method of identifying both the qubit and coupler's spin components is not as universally applicable as more general means such as the Schrieffer-Wolff transformation \cite{bravyi11}, particularly in the strong coupling regime.  However, as we will restrict our analysis to the weak coupling limit, the results of the two methods coincide \cite{quintana17,consani20,khezri21,khezri21_2}.  With these assumptions, the behavior of the physical device can be mapped to the one-dimensional, transverse field Ising spin model.    

\begin{figure*}[t]
\centering
\includegraphics[width = \linewidth]{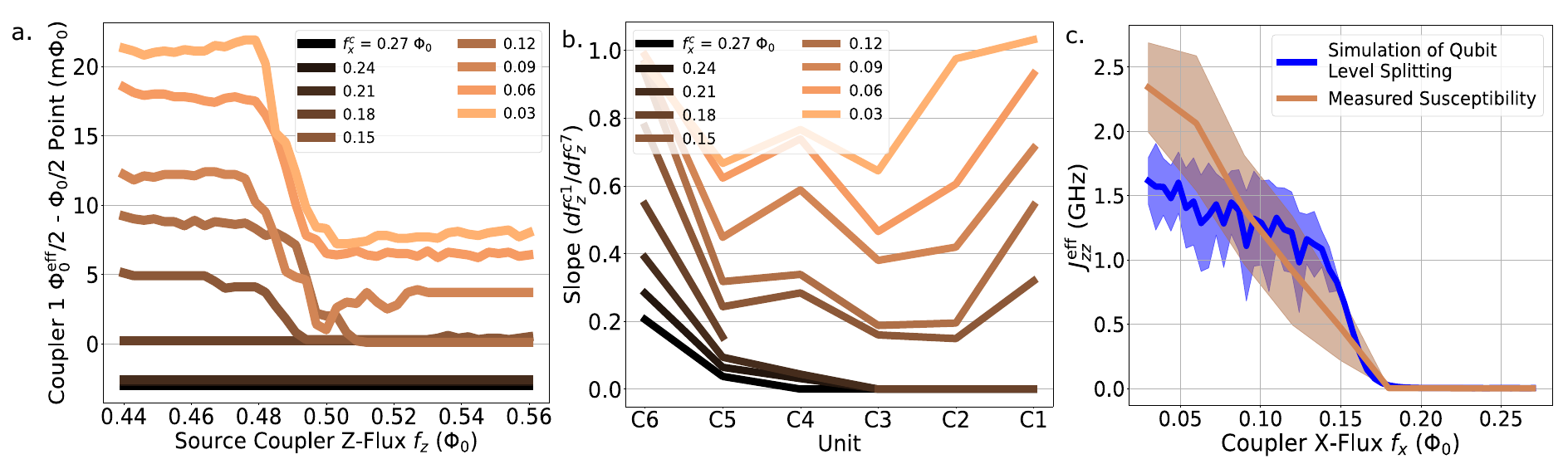}
\caption{\textbf{Cross Chain Susceptibility.}  \textbf{a)}  The effective  one-half magnetic flux quantum points for Coupler 1's main loop as a function of Coupler 7's $f_z$ for different homogeneous coupler $f_x$ settings.  The response becomes non-zero in the vicinity of the couplers' $f_x = 0.15\, \Phi_0$.  The offset from $f_z = \Phi_0/2$ is due to slight mistunings of the couplers' longitudinal fields.  The uncertainty in the extracted values is identical to those in Fig. 3.  \textbf{b)}  The curves in Panel \textbf{(a)}, and the corresponding figures for the other units, were then fit to a sigmoid function.  Displayed here is the midpoint slope extracted from those fits for all units for different homogeneous coupler $f_x$ settings.  The error bars were generated by repeating this procedure many times with small offsets picked from a normal distribution with standard deviation of 1.2 m$\Phi_0$ applied to the nominal effective symmetry point values.  The error bars are the resultant standard deviation of the large array of slopes extracted from the fit sigmoid function.   \textbf{c)}  Using the slopes from Panel \textbf{(b)} for Coupler 1, the effective coupling between the two qubits is calculated via Eqs. (9) and (10) and displayed.  This is compared to $J^{\rm{eff}}_{\rm{q}_1 \rm{q}_2}$ calculated from the splitting of the otherwise degenerate qubit transitions as predicted by full device simulations incorporating reasonable levels of flux noise.  The strong agreement between the two different measures of $J^{\rm{eff}}_{\rm{q}_1 \rm{q}_2}$ indicate that the optimal operating region of the chain is coupler $0.15 \lesssim f_x \leq 0.18 \,\Phi_0$, where there is significant coupling strength and, as shown in Fig. 5, both the detrimental effects of flux noise and the qubit-coupler state mixing is negligible.}
\end{figure*}

Each coupler circuit can be approximately characterized by its susceptibility, $\chi$, which is the change in current induced by a biasing flux.  Assuming the coupler remains in its ground state, this is equivalent to the curvature of the ground state energy with respect to the flux in the coupler's main loop.
\begin{equation}
    \chi = \frac{1}{L_{\rm{eff}}} = \frac{d\langle I^z_{\rm{c}} \rangle}{d\,f_z} \approx 
    \frac{d^2 E^0_{\rm{c}}}{d\,f_z^2}
\end{equation}
In Eq. (7), $\langle I^z_{\rm{c}} \rangle$ represents the ground state expectation value of the current in the coupler's main loop and $E_{\rm{c}}^0$ is the ground state energy of the coupler unit.  The character of these two quantities, $\langle I^z_{\rm{c}} \rangle$ and $E_{\rm{c}}^0$, is determined by $f_x$, the magnetic flux in the coupler's small loop.  The coupler circuit's susceptibility, $\chi$, is optimized as the unit's $\beta_{\rm{c}} \equiv L_{\rm{c}}/L_{\rm{eff}} =  2\pi L_{\rm{c}} I_{\rm{c}}^{(c)}/\Phi_0 \approx 1$, where the coupler's local potential minimum is highly sensitive to biasing flux.  As shown in Fig. 2, this occurs in the same $f_x$ region where $J_{\rm{cc}}/(\Delta_{\rm{c}} / 2) \approx 1$, a design choice made to optimize the generation of long-range correlations across the device.  In addition, the device can be operated in a regime such that the coupler's minimum excitation energy, larger than 5 GHz, is much greater than the temperature of the system, approximately 400 MHz, the strength of the qubit-coupler interaction, which is below 1 GHz, and the typical qubit excitation frequency, approximately 2 GHz.  This ensures the ground state properties of the coupler dictate its behavior, entanglement between qubits mediated by the coupler is supported, and fast (coupler) and slow (qubit) modes can be separated to preserve the qubit subspace.  

The same notions of susceptibility and design constraints can also be applied to a long but finite chain of couplers.  The current induced in coupler $j$ when a flux is applied to coupler $i$ is expressed as $\chi_{\rm{c}_i \rm{c}_j}$, the inter-chain susceptibility.  The design constraints are slightly more involved for the chain when compared to a single coupler.  For example, the length of the chain has a closing effect on the size of the gap as the fundamental mode frequency of the chain decreases with length.  This introduces a trade-off between the physical range of interaction and the need to preserve the excitation energy gap of the chain \cite{meier03}.  

Taking the same concept of single coupler susceptibility to hold for the coupler chain, as well as adhering to the extra constraints introduced by the many-body chain system, it is possible to construct the form of the effective qubit-qubit interaction, Eq. (6), in terms of circuit parameters.  Assuming the coupler gap is much greater than the qubit working frequencies allows one to separate the device spectrum into `slow' qubit-like states and `fast' coupler-like states.  Invoking the Born-Oppenheimer approximation restricts the coupler spectrum to its unperturbed ground state \cite{melanson19}.  Further restricting our analysis to the weak coupling limit, $M_{\rm{qc}}/L_{\rm{c}} \ll 1$, allows us to write the qubit-chain interaction, in general a complicated nonlinear quantity, as an inductive interaction between qubit currents \cite{quintana17,kafri17}, given by
\begin{equation}\label{Jeff}
\begin{split}
    H_{\rm{int}} 
    = \chi_{\rm{c}_1 \rm{c}_7} & (M_{\rm{q}_1 \rm{c}_1} \hat{I}^z_{\rm{q}_1})(M_{\rm{q}_2 \rm{c}_7} \hat{I}^z_{\rm{q}_2}) \\
    = \widetilde{\chi}_{\rm{c}_1 \rm{c}_7}& (J_{\rm{q}_1 \rm{c}_1})(J_{\rm{q}_2 \rm{c}_7})\,\sigma^z_{\rm{q}_1}\sigma^z_{\rm{q}_2}.
\end{split}
\end{equation}
The final line of Eq. (8) connects the circuit model of the coupler chain to the spin chain model by recognizing that the two versions of the susceptibility are related by $\widetilde{\chi}_{\rm{c}_1 \rm{c}_7}\,=\,\frac{\chi_{\rm{c}_1 \rm{c}_7}}{I_{\rm{c}_1}^{\rm{p}} I_{\rm{c}_7}^{\rm{p}}}$.  The symbol $I_{\rm{c}_i}^{\rm{p}}$ refers to the persistent current of the $i^{\rm{th}}$ coupler, which, when operated at $f_z\,=\,\Phi_0/2$, is simply the current dipole moment $\bra{0} \hat{I}^z_{\rm{c}_i} \ket{1}$ \cite{consani20}.

Now that the long-range, effective interaction between the two qubits is expressed in terms of circuit parameters, it is possible to measure the response function, $\chi_{\rm{c}_1 \rm{c}_7}$, of the coupler chain in a quantitative manner.  This task is accomplished by performing two similar measurements.  Firstly, only the behavior of the coupler chain units are considered.  This measurement is performed with both flux qubits placed at a magnetic flux bias operating point where the circuit has its maximum transition frequency which is much greater than the operating frequencies of the remaining chain units.  This decouples the qubits from the coupler chain dynamics and allows the coupler susceptibility to be characterized.  Secondly, with knowledge of the chain susceptibility, the two flux qubits are brought into an interacting flux operating point and cross chain qubit-qubit interactions are demonstrated.  Finally, to quantitatively measure the effective qubit-qubit interaction strength, $J^{\rm{eff}}_{\rm{q}_1 \rm{q}_2}$, we revisit the coupler chain only measurements in more detail to extract the cross chain susceptibility, $\chi_{\rm{c}_1 \rm{c}_7}$.  We find that our measurements of $J^{\rm{eff}}_{\rm{q}_1 \rm{q}_2}$ through susceptibility measurements agree well with full device simulations of the qubit-qubit spectral line splitting.

\section{\label{sec:level1}Results}

The coupler chain device consists of two capacitively shunted, tunable flux qubits and seven tunable rf-SQUIDs, all equipped with individual readout resonators.  Device characterization, circuit parameter extraction, and wiring details are described in the Supplementary Materials \cite{sm}.  The device is fabricated using the architecture described in Ref. \cite{rosenberg17}, and consists of two separate chips – called the qubit layer and the interposer layer.  The interposer layer, seated on the device package's printed circuit board cavity and wirebonded to the exterior control lines, holds the flux bias lines.  The qubit layer, hosting the qubits, couplers, resonators, and co-planar waveguide, is indium bump bonded atop.  The indium bumps provide structural stability, common ground paths between layers, and a conduit for microwave signals originating on the interposer layer, running through the bumps, and continuing on the qubit layer.  This device environment allows greater flexibility than planar devices for distributing flux bias lines, represents a step towards full 3-D integration, and supports an electromagnetic environment suitable for quantum annealing controls \cite{sm}.  

Each unit, qubit or coupler, possesses a meandering resonator terminated in an rf-SQUID for purposes of readout and calibration.  These resonators, when their terminating rf-SQUID is biased to a flux sensitive region, act as magnetic flux detectors, capable of discerning the qubit or coupler unit's persistent current state.  When the terminating rf-SQUID is biased to its flux insensitive operating point, the resonator is exclusively sensitive to the unit's energy level occupation through the resonator-unit dispersive interaction \cite{grover20}.  Being able to operate in these two modes alleviates the need for multiple readout structures, further freeing up space on chip.

Gaining full flux control of a device of this size is a difficult task for a number of reasons.  Complete individual control of each unit requires 27 flux bias control lines corresponding to the 27 Josephson flux loops.  Current in one control line provides magnetic flux for its target Josephson loop but also couples to nearby loops.  Hence, it is necessary to determine the full 27$\times$27 element mutual inductance matrix before one can expect adequate control of this device.  In addition, these inductive elements need to be determined while in the presence of spurious interactions between units.  Strong inter-unit interactions can easily mask the linear line-loop inductive interaction.  In order to address these points, scalable, device independent, automated methods have been developed and implemented to characterize the bias line to circuit flux inductive matrix to within acceptable errors for device control \cite{dai21}.

To explore the behavior of long-range qubit interactions mediated by the coupler chain, it is necessary to characterize the inter-coupler susceptibility, $\chi_{\rm{c}_i \rm{c}_j}$.  To isolate the coupler chain dynamics, we first flux bias the two end qubits to their high frequency, uncoupled state.  Every coupler unit is operated such that its main loop is flux biased at one-half magnetic flux quanta and its small loops are uniformly biased with $f_x^{\rm{c}}$ .  In this configuration, Coupler 7's $f_z$ is swept across its half quanta point for a range of uniformly biased coupler $f_x^{\rm{c}}$ values.  Instead of directly measuring the current response in the target unit, we observe the shift of the target unit's effective main loop half-quanta point \cite{sm}.  As the unit's main loop half-quanta flux operating point corresponds to its minimum transition frequency, the dispersive interaction with the unit's resonator provides an accurate determination of the unit's effective half-quanta point in the presence of strong inter-unit and unit-resonator interactions.  

\begin{figure}
\centering
\includegraphics[width = \linewidth]{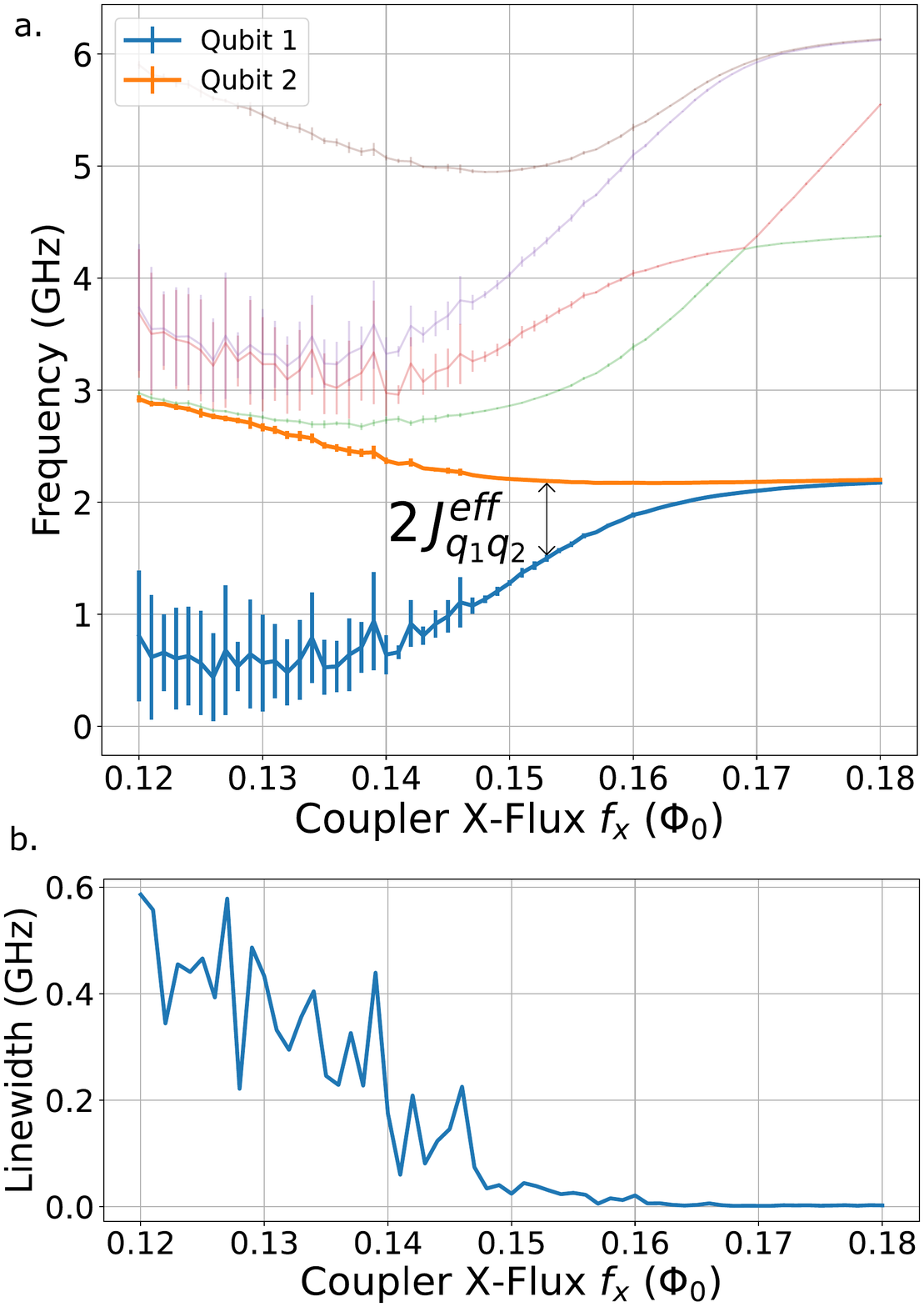}
\caption{\textbf{Effects of Noise.}  \textbf{a)} Full device simulation illustrating the effects of low frequency flux noise on the qubit-like energy levels.  Displayed are the means and standard deviations of the full device energy levels compiling ten separate calculations with the described random flux operating point offsets.  The two lowest energy levels are identified as the two initially degenerate qubit levels, $\ket{+\,-}$ and $\ket{-\,+}$.  In the limit of no coupling, or large coupler $f_x$, they reproduce single qubit behaviour.  As the coupler $f_x$ is lowered and cross chain coupling becomes significant, the previously degenerate qubit levels, shown in bold, split in the presence of cross chain coupling.  Also shown, in lightly faded color, are the few next higher energy levels of the full device consisting of a mixture of coupler and higher qubit levels.  Note that the coupler chain and the qubit levels have become comparable in frequency by coupler $f_x \lesssim 0.15 \,\Phi_0$.  In this region, what were initially qubit-like energy levels, are now dressed by the coupler levels and the splitting of these two energy states no longer represents the effective qubit-qubit interaction strength \cite{dmtnote}.  \textbf{b)} Shown is the linewidth of the lower qubit-like level, calculated as the standard deviation of the transition frequency over multiple simulation runs in the presence of realistic flux noise.  There is a region, $f_x = 0.15-0.18 \, \Phi_0$, where significant cross chain coupling is present but where the effects of flux noise do not significantly broaden the lower qubit-like level's linewidth.  For $f_x < 0.15\,\Phi_0$, the calculated linewidth suggests the coherence times of the qubit have deteriorated to the nanosecond scale.}
\end{figure}

The magnitude of the induced fluxes for each coupler unit are displayed in Fig. 3.  Recall that $f_x^{\rm{c}}$ simultaneously controls the unit's transverse field, $\Delta_{\rm{c}}$, in the Ising spin model picture as well as the magnitude of the unit's persistent current, thus the longitudinal coupling strength, $J_{\rm{cc}} = M_{\rm{cc}}I^z_{\rm{c}_i}I^z_{\rm{c}_{i+1}} $\cite{sm}.  For larger values of $f_x^{\rm{c}}$, the flux propagation signal attenuates over short length scales, up to a few coupler units.  As shown in Fig. 2b, this is the regime where the transverse field dominates and the chain system is in its paramagnetic state.  For $f_x^{\rm{c}} \leq 0.18\,\Phi_0$, long-range correlations are supported across the entire chain.  This is where the critical region of the underlying Ising spin model is expected to be located.

The critical region of the Ising spin model is determined by single coupler properties.  To further validate our results, full device simulations of the experimental protocol were performed.  To perform simulations of a device with such a large Hilbert space, a hierarchical scheme is employed \cite{kerman20_2}.  First, the low energy spectrum, eigenstates, and other operator eigenvalues are computed for individual units.  These eigenstates then form the initial basis for computing the energy spectrum of two and three unit systems comprising subsections of the full device.  Finally, the subsections are appropriately coupled together and the full device energy spectrum, eigenstates, and relevant operator eigenvalues are calculated.  At each step of this procedure, necessary mode occupation numbers are included in the calculation such that the low-lying energy and other operator spectra are well converged.  Simulating the identical procedure as the measurement protocol allows us to track the target unit's effective main loop half-quanta point yielding results that match well with the experimental outcome \cite{sm}. 

With the derived information on the chain susceptibility, we now place the two end qubits at flux operating points where it is possible to couple to the adjacent chain unit.  Qubit 1, the target qubit, is operated such that its main loop is flux biased at one-half magnetic flux quantum and its smaller x-loop is flux biased such that its transverse field has strength $\Delta_{\rm{q}_1} = $ 2.3 GHz, approximately where the qubit's potential becomes double-welled.  Qubit 2, the source qubit, is placed at its minimum $\Delta_{\rm{q}_2} \simeq$ 10 MHz, deep in its double well regime, and its flux bias $f_z$ is swept across its one-half magnetic flux quantum point.  This measurement protocol is repeated for the different coupler chain operating points described in the coupler chain susceptibility experiment.    

The results of the qubit-qubit interaction experiment are shown in Fig. 3.  This figure displays the magnitude of the flux signal propagating along the coupler chain and ultimately into the opposite qubit.  These results agree well with full device simulations \cite{sm} of the equivalent protocol.  As shown in Fig. 3c \& d, long-range, cross chain interactions become supported at approximately $f_x^{\rm{c}} \sim 0.15-0.18\,\Phi_0$ in both the coupler chain susceptibility and long-range qubit interaction experiments.

Furthermore, the full results of the coupler-only susceptibility measurements can be used to predict the strength of the effective qubit coupling, $J^{\rm{eff}}_{\rm{q}_1 \rm{q}_2}$, mediated by the chain.  Equations (9) and (10) show how the measured coupler susceptibility, $\frac{df^z_{\rm{c}_1}}{df^z_{\rm{c}_7}}$, determines the effective qubit interaction strength. 

\begin{align}
J_{\rm{q}_1 \rm{q}_2}^{\rm{eff}}& = \chi_{\rm{c}_1 \rm{c}_7}(M_{\rm{q}_1 \rm{c}_1} I^z_{\rm{q}_1})(M_{\rm{q}_2 \rm{c}_7} I^z_{\rm{q}_2}) \\
        \chi_{\rm{c}_1 \rm{c}_7}& = \frac{d\langle I^z_{\rm{c}_1} \rangle}{df_{\rm{c}_7}^z} = \frac{d\langle I^z_{\rm{c}_1} \rangle}{df_{\rm{c}_1}^z} \frac{df^z_{\rm{c}_1}}{df^z_{\rm{c}_7}}    
\end{align}

Shown in Fig. 4 are the effective one-half magnetic flux quantum points of Coupler 1's main loop as a function of Coupler 7's $f_z$ for various homogeneous coupler unit settings.  As expected, for the flux regime where the transverse fields dominate, Coupler 1's effective half quanta point is unaffected by the $f_z$ of Coupler 7.  As the transverse field is lowered and the coupler-coupler longitudinal coupling strength increases, the effect of Coupler 7's $f_z$ on Coupler 1 becomes more pronounced.  The slope at the center of these curves, $\frac{df^z_{\rm{c}_1}}{df^z_{\rm{c}_7}}$, can then be extracted.  The remaining factors, $I^{z}_{\rm{q}_i}$ and $\frac{d\langle I^z_{\rm{c}_1} \rangle}{df_{\rm{c}_1}^z}$, are determined from single unit simulations \cite{sm}.  

A proper comparison of the qubit interaction strength, $J^{\rm{eff}}_{\rm{q}_1 \rm{q}_2}$, derived from both full device simulations and the measured coupler susceptibility, is necessary to confirm that the coupler susceptibility is a valid measure of qubit coupling strength.  As illustrated in Fig. 5, The effective qubit-qubit interaction strength is derivable from full device simulations by noting the energy level splitting between the two end qubits' previously degenerate energy spectrum.  This is done by first setting both qubits to have the same finite transverse field and zero longitudinal field. As the cross chain coupling is increased by lowering the $f_x^{\rm{c}}$ of all couplers, the initially degenerate qubit states, $\ket{+\,-}$ and $\ket{-\,+}$, develop a splitting that, for weak coupling, is twice the coupling strength 2$J_{\rm{q}_1 \rm{q}_2}^{\rm{eff}}$.  A comparison can then be made between the qubit interaction strength, $J_{\rm{q}_1 \rm{q}_2}^{\rm{eff}}$, as measured in the flux signal propagation experiment, and the qubit level splitting exhibited in full device simulations.    As shown in Fig. 4c, the simulated results find strong agreement with the results of our coupler susceptibility measurements in the weak interaction limit.  The quantitative divergence of these two results is expected in the strong coupling regime, for the coupler flux biases $f_x^{\rm{c}} < 0.15 \,\Phi_0$.  In this case, the weak interaction limit assumed in Eq. (8) breaks down and the effective qubit-qubit interaction can no longer be described as a linear inductive coupling.  Additionally, in the strong coupling regime shown in Fig. 5, the qubit energy levels become dressed by the coupler levels.  Hence, the two lowest energy levels can no longer be identified as purely qubit-like and their spectral distance no longer represents a simple qubit-qubit interaction. 

Low frequency flux noise is expected to play a detrimental role in the coherence preserving properties of this long-range interaction.  As demonstrated, flux signals can be transported and even amplified in certain coupler flux operating regimes.  The question is then, is it possible to identify a flux operating regime for the coupler units where strong, long-range coupling is present, yet the detrimental effects of flux noise are not amplified across the device?  

To answer this, full device simulations were performed with realistic values of low frequency flux noise.  Flux noise has been measured, across many different platforms and frequencies \cite{yoshihara06, yan16, braumuller20}, to be approximately $1/f^\alpha$ in nature, $\alpha\sim0.9$, with magnitude $1-5\, \mu\Phi_o/\sqrt{\rm{Hz}}$.  For moderate frequency measurements, the effect of this low frequency noise is to effectively add a small random flux offset to the flux operating point of the measurement.  This small random flux offset is sampled from a Gaussian distribution whose standard deviation is determined by integrating the noise spectrum over the appropriate frequency range, from measurement repetition rate to pertinent experimental frequencies, as well as accounting for the circuit geometry.  This amounts to a typical random flux offset in the tens of $\mu\Phi_o$. 

Simulations of this type were performed repeatedly to determine the behavior of the device energy level structure in the presence of low frequency flux noise.  As shown in Fig. 5, the energy spectrum of the device is highly susceptible to flux noise in its deeply coupled state.  Significant line broadening occurs for the uniformly tuned coupler $f_x^{\rm{c}}$ between 0 - 0.15 $\Phi_0$.  This allows us to identify a region of flux operation, coupler $f_x^{\rm{c}}$ from 0.15 to 0.18 $\Phi_0$ where significant long range interactions are present yet the detrimental effects of flux noise are still minimal.  

\section{\label{sec:level1}Discussion}

Quantum annealing processors stand to benefit from higher dimensional qubit networks, expanded connectivity, and improved qubit coherence.  Accomplishing this will require long-range qubit interactions that do not degrade qubit behavior.  The use of spin chains as a quantum bus is a promising venue for this.  Presented here is a preliminary step in this direction in the context of a superconducting Josephson system.  To build on this idea, there have been proposals to generalize one dimensional spin chains, capable of entangling end qubits, to both paramagetic trees \cite{weber18,kerman18} and two dimensional spin networks capable of providing entanglement amongst a perimeter of qubits \cite{zippilli13,kerman18,kerman20}.  Such architectures could provide a scalable, coherent quantum annealing device with high graph dimensionality. 

Susceptibility measurements in quantum systems, such as those performed in this study, have been considered as a possible measure of the system's entanglement \cite{hauke16}.  The susceptibility experiment's close agreement with full device simulations, which also demonstrate qubit energy level splitting in the presence of expected noise levels, suggest the qubits can be prepared in an entangled doublet state.  In this view, the susceptibility measurements presented here are a consequence and valid measure of a coherent long-range qubit interaction.  This view, however, needs further experimental validation.  Future experimental work will address measurement of the coherent coupling enabled by this method using spectroscopic characterization as well as adiabatic transfer protocols.  Additionally, the detection of entanglement can be augmented by measuring other observables and witnesses for interacting quantum spin systems \cite{wu05}. 

In closing, we have demonstrated long-range interactions in a superconducting Josephson spin bus by probing the device's response function.  Simulations of the device, which agree well with measured quantities, predict significant long-range interaction simultaneous with satisfactory qubit coherence.  This device has immediate application in near-term quantum annealing devices where both long-range and coherent qubit couplings are necessary for quantum computation speedup.  

\section{Methods}

See Supplementary Material.

\section{Data Availability}

The data that support the findings of this study are available from the corresponding author upon reasonable request.

\section{Competing Interests}

The authors declare no competing interests.

\section{Author Contributions}

D.M.T., A.J.M, and R.T. performed the measurements.  D.M.T., X.D., A.J.M., J.A.G, S.M.D., and J.I.B built the code base supporting the experiment.  A.J.M., D.M., and S.N. designed the device.  D.M.T., D.M., M.A.Y., Y.T., S.B., and R.Y. provided analysis of results.  R.D, D.K.K., A.J.M., B.M.N., and J.L.Y. fabricated the device.  S.J.W., A.J.K., E.M., D.A.L, and A.L. supervised the project.

\section{ACKNOWLEDGMENTS}

We wish to thank the members of the Quantum Enhanced Optimization/Quantum Annealing Feasibility Study collaboration for various contributions that impacted this research.  In particular, we thank Lorenzo Campos Venuti and Cyrus F. Hirjibehedin for helpful discussions on spin chains and device design and Ken Zick and David Ferguson for various discussions on experiments.  We gratefully acknowledge the MIT Lincoln Laboratory design, fabrication, packaging, and testing personnel for valuable technical assistance.  

The research is based upon work (partially) supported by the Office of the Director of National Intelligence (ODNI), Intelligence Advanced Research Projects Activity (IARPA) and the Defense Advanced Research Projects Agency (DARPA), via the U.S. Army Research Office contract W911NF-17-C-0050. The views and conclusions contained herein are those of the authors and should not be interpreted as necessarily representing the official policies or endorsements, either expressed or implied, of the ODNI, IARPA, DARPA, or the U.S. Government. The U.S. Government is authorized to reproduce and distribute reprints for Governmental purposes notwithstanding any copyright annotation thereon.

Daniel M. Tennant acknowledges the support of the U.S. Department of Energy by Lawrence Livermore National Laboratory under Contract DE-AC52-07NA27344 while preparing this document.

\bibliography{MainText}

\begin{thebibliography}{76}
\bibitem [{qpt()}]{qpt_note}%
  \BibitemOpen
  \href@noop {} {}\bibinfo {note} {More precisely, these parameters are where
  the quantumphase transition happens for an infinite system.}\BibitemShut
  {Stop}%
\bibitem [{dmt()}]{dmtnote}%
  \BibitemOpen
  \href@noop {} {}\bibinfo {note} {As can be seen in Fig. 5a, a coupler chain
  state anti-crosses with the qubit state $\ket{\uparrow \uparrow}$ at
  approximately coupler $f_x$ = 0.17 m$\Phi_0$. While this does appear to
  violate our previous assumptions concerning the separation of coupler chain
  and qubit operating frequencies, for adiabatic annealing protocols the
  $\ket{\uparrow \uparrow}$ state should rarely be accessed. Additionally this
  property can be improved upon in future design iterations}\BibitemShut
  {NoStop}%
\end{thebibliography}

\onecolumngrid

\listoffigures

\end{document}


\title{\label{sec:level1}Supplementary Materials for Demonstration of long-range correlations via susceptibility measurements in a one-dimensional superconducting Josephson spin chain}

\author{D. M. Tennant}
\thanks{Present Address: Lawrence Livermore National Laboratory, Livermore, California 94550, USA}
\email{tennant3@llnl.gov}
\affiliation{Institute for Quantum Computing, University of Waterloo, Waterloo, ON, Canada N2L 3G1}
\affiliation{Department of Physics and Astronomy, University of Waterloo, Waterloo, ON, Canada N2L 3G1}


\author{X. Dai}
\affiliation{Institute for Quantum Computing, University of Waterloo, Waterloo, ON, Canada N2L 3G1}
\affiliation{Department of Physics and Astronomy, University of Waterloo, Waterloo, ON, Canada N2L 3G1}

\author{A. J. Martinez}
\affiliation{Institute for Quantum Computing, University of Waterloo, Waterloo, ON, Canada N2L 3G1}
\affiliation{Department of Physics and Astronomy, University of Waterloo, Waterloo, ON, Canada N2L 3G1}

\author{R. Trappen}
\affiliation{Institute for Quantum Computing, University of Waterloo, Waterloo, ON, Canada N2L 3G1}
\affiliation{Department of Physics and Astronomy, University of Waterloo, Waterloo, ON, Canada N2L 3G1}

\author{D. Melanson}
\affiliation{Institute for Quantum Computing, University of Waterloo, Waterloo, ON, Canada N2L 3G1}
\affiliation{Department of Physics and Astronomy, University of Waterloo, Waterloo, ON, Canada N2L 3G1}

\author{M. A. Yurtalan}
\affiliation{Institute for Quantum Computing, University of Waterloo, Waterloo, ON, Canada N2L 3G1}
\affiliation{Department of Physics and Astronomy, University of Waterloo, Waterloo, ON, Canada N2L 3G1}

\author{Y. Tang}
\affiliation{Institute for Quantum Computing, University of Waterloo, Waterloo, ON, Canada N2L 3G1}
\affiliation{Department of Physics and Astronomy, University of Waterloo, Waterloo, ON, Canada N2L 3G1}

\author{S. Bedkihal}
\affiliation{Institute for Quantum Computing, University of Waterloo, Waterloo, ON, Canada N2L 3G1}
\affiliation{Department of Physics and Astronomy, University of Waterloo, Waterloo, ON, Canada N2L 3G1}

\author{R. Yang}
\affiliation{Institute for Quantum Computing, University of Waterloo, Waterloo, ON, Canada N2L 3G1}
\affiliation{Department of Physics and Astronomy, University of Waterloo, Waterloo, ON, Canada N2L 3G1}

\author{S. Novikov}
\affiliation{Northrop Grumman Corporation, Linthicum, Maryland 21090, USA}

\author{J. A. Grover}
\affiliation{Northrop Grumman Corporation, Linthicum, Maryland 21090, USA}

\author{S. M. Disseler}
\affiliation{Northrop Grumman Corporation, Linthicum, Maryland 21090, USA}

\author{J. I. Basham}
\affiliation{Northrop Grumman Corporation, Linthicum, Maryland 21090, USA}

\author{R. Das}
\affiliation{MIT Lincoln Laboratory, Lexington, Massachusetts 02421, USA}

\author{D. K. Kim}
\affiliation{MIT Lincoln Laboratory, Lexington, Massachusetts 02421, USA}

\author{A. J. Melville}
\affiliation{MIT Lincoln Laboratory, Lexington, Massachusetts 02421, USA}

\author{B. M. Niedzielski}
\affiliation{MIT Lincoln Laboratory, Lexington, Massachusetts 02421, USA}

\author{S. J. Weber}
\affiliation{MIT Lincoln Laboratory, Lexington, Massachusetts 02421, USA}

\author{J. L. Yoder}
\affiliation{MIT Lincoln Laboratory, Lexington, Massachusetts 02421, USA}

\author{A. J. Kerman}
\affiliation{MIT Lincoln Laboratory, Lexington, Massachusetts 02421, USA}

\author{E. Mozgunov}
\affiliation{Center for Quantum Information Science \& Technology, University of Southern California, Los Angeles, California 90089, USA}

\author{D. A. Lidar}
\affiliation{Center for Quantum Information Science \& Technology, University of Southern California, Los Angeles, California 90089, USA}
\affiliation{Departments of Electrical \& Computer Engineering, Chemistry, and Physics, University of Southern California, Los Angeles, California 90089, USA}

\author{A. Lupascu}
\affiliation{Institute for Quantum Computing, University of Waterloo, Waterloo, ON, Canada N2L 3G1}
\affiliation{Department of Physics and Astronomy, University of Waterloo, Waterloo, ON, Canada N2L 3G1}
\affiliation{Waterloo Institute for Nanotechnology, University of Waterloo, Waterloo, ON, Canada N2L 3G1}



\date{\today}

\maketitle

\beginsupplement

\tableofcontents

\newpage

\section{\label{sec:level1}Spin Chain Model}

For the derivation of the effective qubit-qubit interaction strength, we assume that the coupler chain is operated in its paramagnetic regime such that its ground state is non-degenerate and the energy gap to its first excited state is sufficiently larger than the temperature of the system, the qubit-coupler interaction strength, and the qubit operating frequencies.  As the chain approaches its critical region, where we expect long-range correlations to emerge, the gap does become smaller.  But, as shown in Fig. 5 of the main text, the gap still exceeds these other energy scales.

Treating the qubit-chain interaction as a small perturbation on the coupler chain system allows us to write
\begin{equation}
    H_{\mathrm{eff}} \simeq \sum\limits_{i=1}^2 \overline{H}_{q_i} + J^{\mathrm{eff}}_{q1q2}\sigma^z_{q_1}\sigma^z_{q_2}.
\end{equation}
First order perturbation expansion shifts the longitudinal field of the qubits,
\begin{equation}
    \overline{H}_{q_i} = \frac{\overline{\epsilon}_{q_i}}{2}\sigma^z_{q_i} + \frac{\Delta_{q_i}}{2}\sigma^x_{q_i},
\end{equation}

\begin{equation}
    \overline{\epsilon}_{q_i} = \epsilon_{q_i} + J_{q_ic_{\mathrm{adj}}} \bra {0_c} \sigma^z_{c_{\mathrm{adj}}} \ket{0_c},
\end{equation}
where $J_{q_ic_{\mathrm{adj}}}$ = $J_{q_1 c_1}$ or $J_{q_2 c_7}$.
The effective qubit-qubit interaction, mediated by virtual excitations of the chain out of its ground state, can be calculated to second order in perturbation theory as 
\begin{equation}
    \begin{split}
   J^{\mathrm{eff}}_{q_1 q_2} &= - J_{q_1 c_1} J_{q_2 c_7} \sum\limits_{n_c=1}^{2^7-1}\frac{\bra {0_c} \sigma^z_{c_1} \ket{n_c} \bra {n_c} \sigma^z_{c_7} \ket{0_c}}{\omega_{0,n_c}} \\
    &\simeq - J_{q_1 c_1} J_{q_2 c_7} \sum\limits_{n_c=1}^{2^7-1}\frac{\bra {0_c} \sigma^z_{c_1} \ket{n_c} \bra {n_c} \sigma^z_{c_7} \ket{0_c}}{\Omega},
    \end{split}
\end{equation}
where $\Omega\equiv\omega_{01}$, the energy gap between the ground and first excited state.  In Eq. (4) it is recognized that there is a band of $N=7$ approximately degenerate excited states with gap $\Omega$ and finite matrix elements from the ground state.  The higher coupler chain levels past this have, at best, exponentially (in energy difference) small $\sigma^z_{c_i}$ matrix elements from the ground state and can be neglected.  Now the outer product of states, $\sum\limits_{n_c=1}^{2^7-1} \ket{n_c} \bra {n_c}$ can be written as $\mathbb{1} - \ket{0_c} \bra {0_c}$.  The effective coupling strength then reduces to 
\begin{equation}
    J^{\rm{eff}}_{q_1 q_2} = \frac{J_{q_1 c_1} J_{q_2 c_7}}{\Omega} (\bra {0_c} \sigma^z_{c_1} \ket{0_c} \bra {0_c} \sigma^z_{c_7} \ket{0_c} - \bra {0_c} \sigma^z_{c_1} \sigma^z_{c_7} \ket{0_c}).
\end{equation}

\newpage

\section{\label{sec:level1}Circuit Model}

The Coupler Chain device consists of 7 tunable rf-SQUIDS coupled through their z-loop mutual inductance with the two end couplers coupled mutually to two capacitively shunted flux qubits.  Shown in Fig. 1 are schematics of the qubit and coupler design.  The qubit and coupler are divided into four and three floating islands respectively.  Design work utilized Ansys Maxwell finite element electromagnetic simulations supplemented with simulations run in Sonnet to correctly account for higher frequency inductive effects.  Effort was made to reproduce effective circuit parameters found in \cite{yan16,weber17}.  The results of these simulations are displayed in Table 1.  

\begin{figure}[h]
\centering
\includegraphics[width = \linewidth]{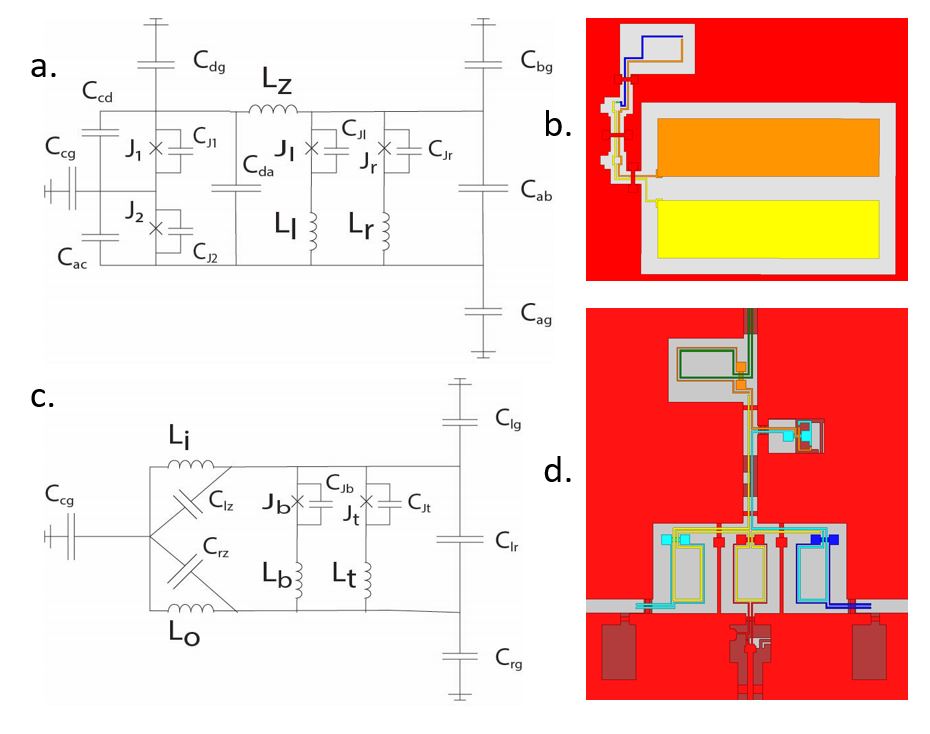}
\caption{\textbf{Qubit \& Coupler Details.}  \textbf{a) \& b)}  Shown is the schematic and layout of the capacitively shunted flux qubit.  \textbf{c) \& d)}  Shown is the schematic and layout of the coupler unit.}
\end{figure}

\begin{table}[h]
\begin{tabular}{ |c|c|c|c|  }
\hline
\multicolumn{2}{|c|}{Qubit Properties} & \multicolumn{2}{c|}{Coupler Properties}\\ 
\hline
$I_c^L$ & 210 nA & $I_c$ & 240 nA\\
$I_c^S$ & 90 nA & & \\
$C_{ga}$ & 97.4 fF & $C_{gr}$ & 32.7 fF \\
$C_{gb}$ & 62.0 fF & $C_{gl}$ & 25.8 fF \\
$C_{gc}$ & 0.17 fF & $C_{gz}$ & 49.5 fF\\
$C_{gd}$ & 30.4 fF & & \\
$C_{ab}$ & 9.81 fF & $C_{rl}$ & 6.54 fF\\
$C_{ac}$ & 0.032 fF & $C_{rz}$ & 7.22 fF\\
$C_{ad}$ & 2.49 fF & $C_{lz}$ & 7.19 fF\\
$C_{cd}$ & 0.29 fF & & \\
$L_{l}$ = $L_{r}$ & 78 pH & $L_x$ & 31.4 pH\\
$L_z$ & 690 pH & $L_{i}$ = $L_{o}$ & 378 pH\\
$M_{qc}$ & 62.6 pH & $M_{cc}$ & 64.2 pH \\
\hline
\end{tabular}
\caption{\textbf{Qubit \& Coupler Circuit Parameters.}  Capacitance and Inductance values extracted respectively from Ansys Maxwell and Sonnet circuit simulations.}
\end{table}

\newpage

\section{\label{sec:level1}Simulation Framework}

Upon attaining all lumped element circuit parameters from the classical electromagnetic Ansys Maxwell and Sonnet simulations, quantum simulations of the lumped element single qubit and coupler units and full device were performed using the MIT-Lincoln Laboratory developed JJSim quantum circuit simulation package \cite{kerman20_2}.  With this software package, energy spectra and operator matrix elements can be extracted.  Single unit simulations are performed by identifying internal modes and evaluating the circuit Hamiltonian in the harmonic level basis for high enough levels of internal modes such that the operator expectation values in the low energy qubit subspace converge.

For full device simulations, individual units were grouped and coupled to one another in a hierarchical structure.  The end qubits remain their own two subgroups.  The coupler chain was partitioned into groups of couplers (1,2), (3,4,5), and (6,7).  These individual groups then underwent convergence tests exploring how many group energy levels need to be included in simulations such that the low energy operator expectation values converge.  Each of the 5 groups were then coupled together for the full device simulations.  Again, convergence tests were performed at this final level.  Finally, to validate our simulated results with our agreed upon energy level structure, energy levels at lower levels of the hierarchy were varied to double check the validity of the simulated results.

The JJsim lumped element simulations were then incorporated into the full simulation workflow, from classical electromagnetic simulations of the physical device to energy spectrum simulations of the derived lumped element circuits, to optimize the device design.  Once the device design was finalized, an additional simulation platform calling JJSim was then developed.  This enables an efficient means of adding variation in circuit parameter and flux value assignment in order to account for low frequency flux noise and fabrication uncertainties, as well as visual result presentation. 

\newpage

\section{\label{sec:level1}Device details}

This device consists of two highly coherent superconducting tunable capacitively shunted flux qubits and seven similarly designed tunable rf-SQUIDs all equipped with individual readout.  These are aluminum devices with Al/AlO$_x$/Al junctions.  

\begin{figure}[h]
\centering
\includegraphics[width = \linewidth]{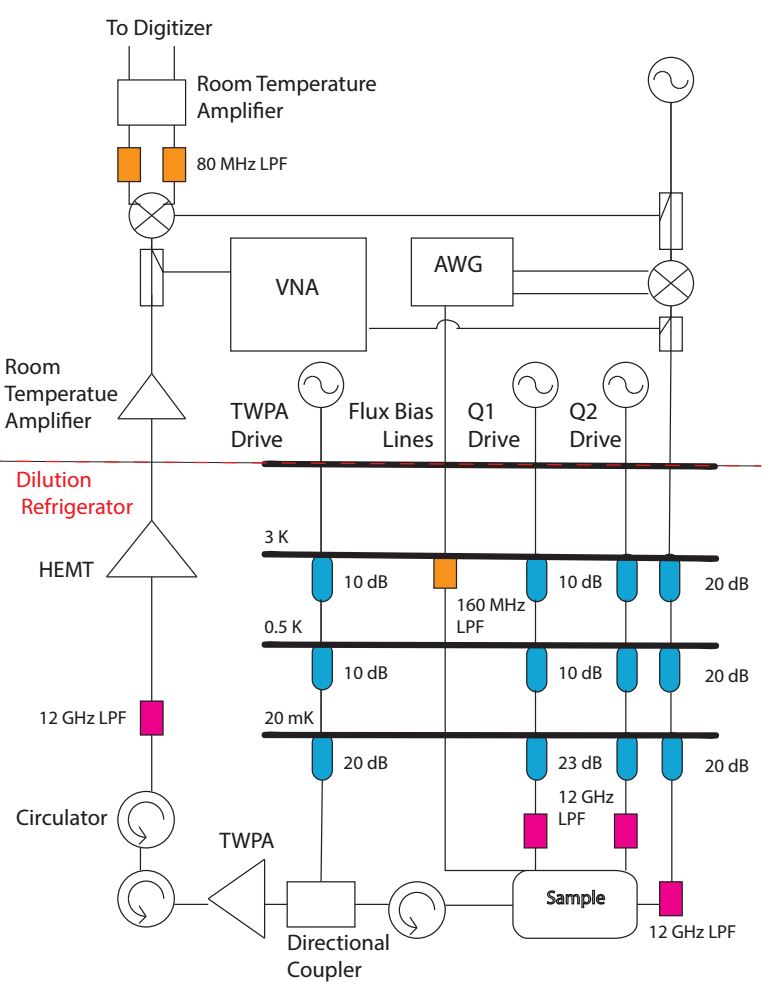}
\caption{\textbf{Dilution Refrigerator Wiring.}  Shown is the wiring schematic used in the coupler chain experiment.}
\end{figure}

This device is hosted in a two-tier environment, illustrated in Fig. 3, similar to the one described in \cite{rosenberg17}.  The interposer tier, sitting in the device package and wirebonded to a PCB which routes signals to the exterior control lines, holds the flux bias lines.  The qubit tier, hosting the qubits, couplers, resonators, and co-planar waveguide, is Indium bump bonded atop.  Both tiers are built out of a thick silicon substrate layer topped by the high quality Aluminum film.  This device environment allows greater flexibility in distributing flux bias lines, represents a step towards full 3-D integration, and supports an electromagnetic environment suitable for quantum annealing controls.  The device package is gold-plated copper with aluminum wirebonds and installed within a dilution refrigerator with base temperature of $\sim$ 20 mK.

\begin{figure}[h]
\centering
\includegraphics[width=\linewidth]{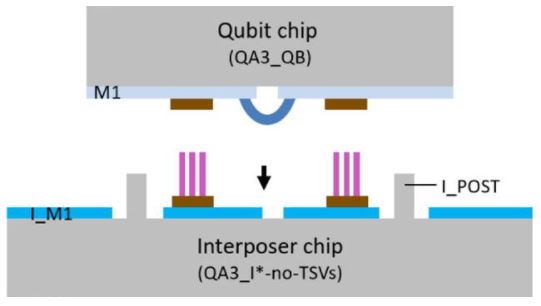}
\caption{\textbf{The Two-Tier Qubit Environment.}  The interposer tier, comprising a Silicon substrate covered in high quality Aluminum, houses the interposer chip.  Also shown in the figure are the Indium bumps and Silicon posts, included for structural support.  The qubit tier, housing the qubit chip, is fabricated with the same materials and fitted atop the interposer tier following fabrication.}

\end{figure}

\begin{figure}[h]
\centering
\includegraphics[width = \linewidth]{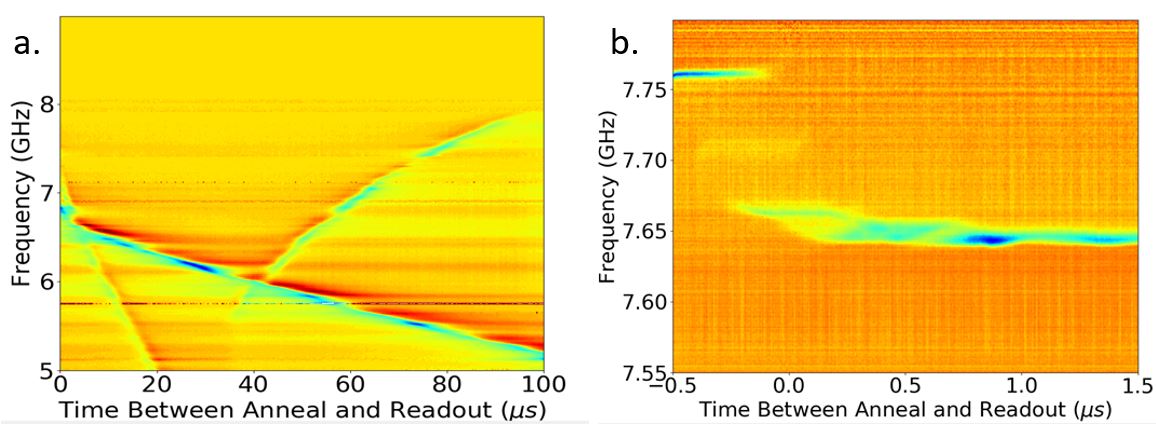}
\caption{\textbf{Flux Transient Measurements.}  The dynamic resonator response is measured in both a single tier \textbf{(a)} and two-stack \textbf{(b)} device environment following a typical annealing-type flux pulse.  The single tier devices have electromagnetic ring-down times of 100$s$ of $\mu$s where as the double stack environment is shielded from this effect, settling into its equilibrium flux within one $\mu$s.}
\end{figure}

\begin{figure*}[h]
\centering
\includegraphics[width = \linewidth]{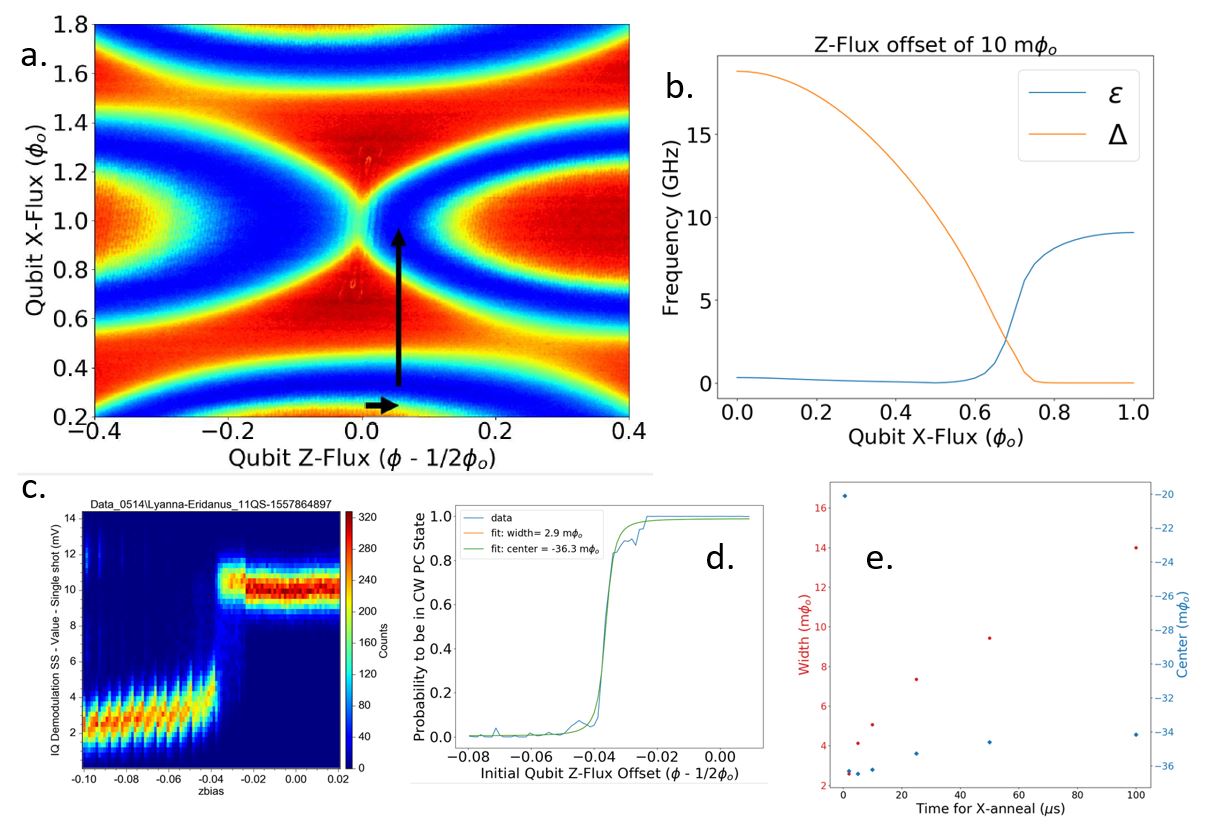}
\caption{\textbf{Single Qubit Annealing Experiments.}  \textbf{a)}  Shown is a 'topographical' map of the qubit in flux space.  The arrows indicate the path in flux space for a typical annealing experiment.  First, a small offset in $f_z$ is applied.  Then the change in $f_x$ is applied.  \textbf{b)}  Shown are the dynamics of the Ising coefficients during the $f_x$ journey.  Initially the qubit is dominated by the transverse field.  As the $f_x$ journey continues the transverse field turns off while the longitudinal field increases.  There is a minimum gap at $\simeq 0.68 \, \Phi_o$ where these two energy scales are equal and the nature of the qubit ground state transitions between the $\hat{\sigma}_x$ and $\hat{\sigma}_z$ basis.  \textbf{c)}  Single shot data showing the histogrammed output voltage of 10,000 runs as a function of initial qubit $f_z$.  The bi-modal distribution of output voltages correspond to the measuring frequency being on-resonance with the qubit in its `left' circulating current state and then off-resonance with the qubit in its `right' circulating current state.  \textbf{d)}  The output voltages are then thresholded and the ground state probability is displayed as a function of qubit $f_z$.  \textbf{e)}  The center $f_z$ and transition widths are plotted as a function of $f_x$ journey time.  The shift in center is a reflection of the qubit x-loop junction asymmetry.  For a range of longer annealing times, the system thermally exits its ground state at different points along the anneal, shifting the effective $f_z$ symmetry point.  These annealing pulses were conducting through low frequency lines, acting as low pass filters of approximately 20 MHz.  The optimal annealing time of $\sim$ 1 $\mu$s results in a width of $\sim$ 3 m$\Phi_0$, similar to previously reported time and widths on these flux bias lines \cite{grover20}.}
\end{figure*}

Single unit coherence and annealing measurements have been performed in the two-stack environment.  Coherence times measured match well with measurements taken on similar qubits in a single planar tier \cite{novikov18}.  Single qubit annealing experiments, as shown in Fig. 5, were also performed yielding comparable transition widths as those performed on planar devices \cite{khezri21}.  

In addition to these encouraging results, the two-stack qubit environment also provides a `clean' electromagnetic environment.  Initial annealing experiments on single-tier qubit environments were hindered by long electromagnetic ring-down times following annealing flux control pulses.  As shown in Fig. 4, this could be measured by tracking the flux-dependent frequency of the readout resonator.  Settling times for the planar geometry devices could reach 100s of $\mu$s.  Repeating these measurements in the two-stack environment yields settling times of approximately a microsecond.  Our initial investigations into this effect broadly agree with the results of \cite{foxen18}, i.e., that the long ring down times can be attributed to charge redistribution of the normal metal ground plane but that the superconducting ground planes on both tiers of the two-stack effectively shield the Josephson elements from this effect.

Each qubit/coupler unit is equipped with its readout resonator.  These resonators are quarter wavelength meandering resonators terminated with an rf-SQUID.  To model this component we consider a quarter wave transmission line resonator \cite{pozar11} terminated in a variable inductor whose value is dictated by the rf-SQUID properties \cite{tinkham04}.  Bare resonator frequencies came out slightly higher than design values by 0.5-2\%.  To investigate this discrepancy, three parameter fits, involving resonator length, junction critical current, and geometrical inductance in one case and a common phase velocity, junction critical current, and geometrical inductance in another were applied to the resonator spectra.  In the first case, the resonator lengths came out consistently ~2\% lower than design values, while in the second case, the phase velocity turned out to be ~2.4\% larger than design.  Both cases showed a similar spread in values of critical currents and geometric inductances.

Qubit 2's circuit parameter values were extracted by matching spectroscopy measurements to single unit circuit simulations described above.  Qubit 1 suffered an unresponsive capacitive drive line during the experimental run.  Displayed in Fig. 6, the data is well fit to the simulation and extracted values agree well with design.

\begin{figure*}[h]
\centering
\includegraphics[width = \linewidth]{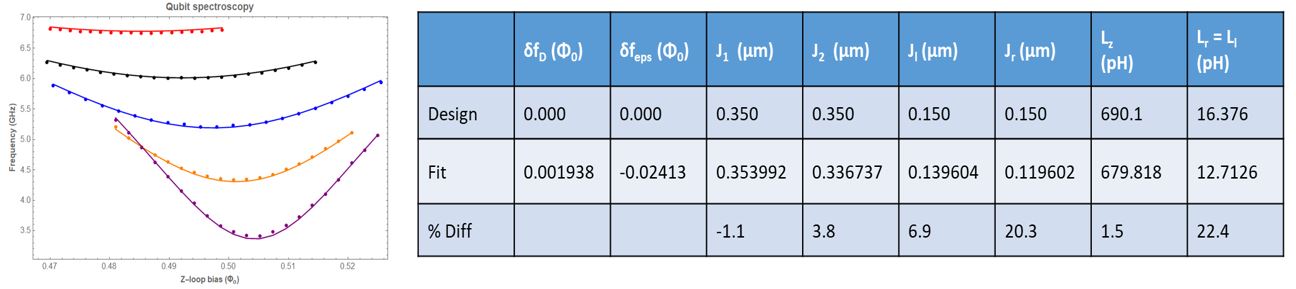}
\caption{\textbf{Qubit 2 Parameter Extraction.}  Two-tone spectroscopy measurements were taken for Qubit 2 for a range of $f_x$ values.  Using a rejection sampling method, multi-variable fits were performed on various circuit parameter.  As shown in the table, all parameters match well with design values.}
\end{figure*}

From coupler `topographical' scans, i.e., probing the coupler resonator at a fixed frequency while varying the coupler's two flux bias currents as shown in Fig. 7, we are able to extract couplers' x-loop junction asymmetry, $d = \frac{J_{x_1} - J_{x_2}}{J_{x_1} + J_{x_2}}$.  Even small asymmetries of a few percent, due to fabrication imperfections, can appreciably shift the z-symmetry point of the qubit/coupler unit.

\begin{figure*}[h]
\centering
\includegraphics[width = \linewidth]{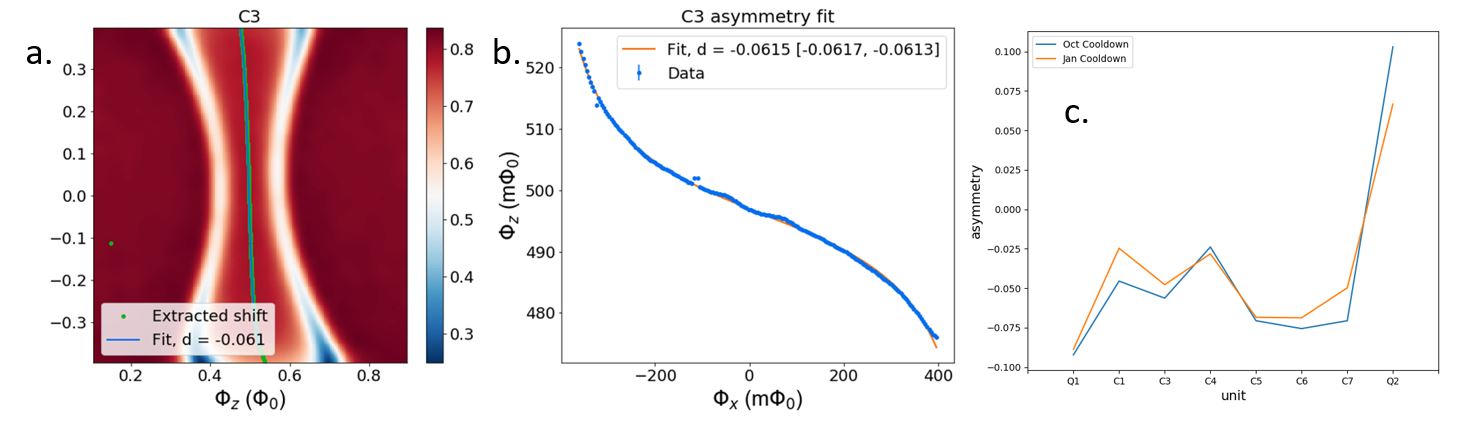}
\caption{\textbf{Coupler X-Loop Junction Asymmetry.}  \textbf{a)} A `topographical' scan of Coupler 3.  By performing inversion symmetry analysis on these data sets \cite{dai21}, it is possible to extract the effective $\Phi_o/2$ points.  \textbf{b)}  These points are well fit by the x-loop junction asymmetry model.  \textbf{c)}  Shown are the asymmetry parameters for the coupler units measured approximately 4 months apart during different dilution refrigerator cooldowns.}
\end{figure*}

\newpage

\section{\label{sec:level1}Experimental Methods}

Both the coupler-only susceptibility measurement and full device qubit interaction demonstration experiments were conducted in similar fashion.  Either Coupler 7, for the coupler-only, or Qubit 2, in the full device, acted as the source unit.  For uniformly tuned coupler units, all couplers at their z-symmetry point and homogeneously tuned x-flux settings, the source unit's z-flux is swept across its z-symmetry point, causing the source unit to transition from its `left' to `right' circulating current state.  This acts to shift the effective z-symmetry point of the remaining units in a manner dependent on the coupler units x-flux.

The response of the target units is measured by observing the dispersive shift in the target units' resonator.  The target units' resonators are maintained at zero flux or their high frequency point.  The dispersive interaction between the unit and resonator can be used to ascertain the unit's z-symmetry point.  So, for each source unit's z-flux setting, the target unit's resonator is probed over a range of frequencies for a range of target unit's z-flux settings.  In this way, the target unit's effective z-symmetry point can be tracked as a function of the source unit's z-flux setting.  

\begin{figure*}[h]
\centering
\includegraphics[width = \linewidth]{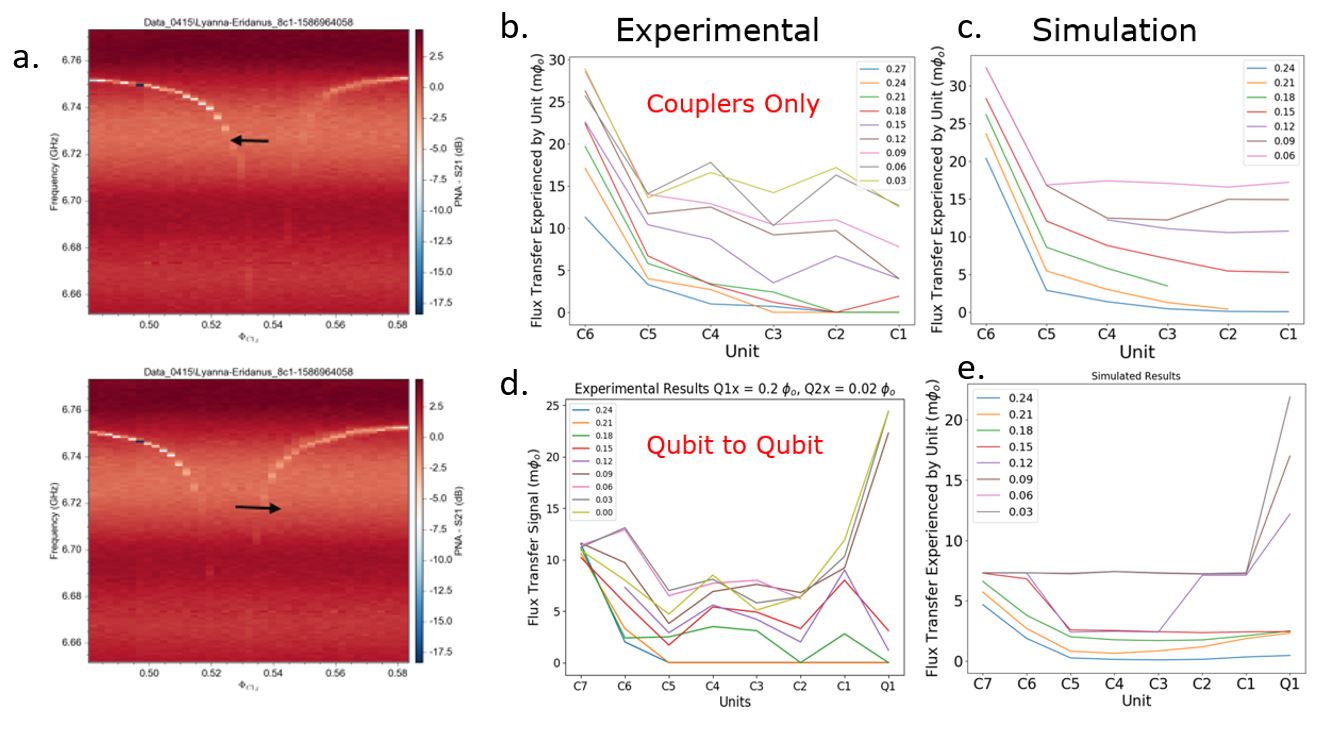}
\caption{\textbf{Flux Propagation Signal Methods.}  \textbf{a)}  Shown are the target unit's resonator for the source unit being in its 'Left' and 'Right' circulating current state.  The resonator is maintained at $f_r = 0$ while the unit $f_z$ is swept.  The dispersive unit-resonator interaction is apparent and can be used to ascertain the unit's $\Phi_o/2$ point.  Automated inversion symmetry methods were employed to extract the $\Phi_o/2$ point.  \textbf{b) \& c)}  The experimental and simulated results showing the magnitude of the coupler-only flux signal at the different units for various homogenously tuned coupler $f_x$ settings.  \textbf{d) \& e)}  The experimental and simulated results showing the magnitude of the full device flux signal at the different units for various homogenously tuned coupler $f_x$ settings.  Both experiments show a strong agreement with simulated results.}
\end{figure*}

This procedure was applied iteratively starting with the unit adjacent to the source unit, and then continued down the chain.  Applying this process iteratively also allowed us to fine-tune the target units' z-flux.  As the coupler units' x-flux is lowered and the inter-coupler interactions become stronger, slight mis-tunings of the unit's z-flux away from their symmetry point can shift the effective symmetry point of nearby units.  Applying the measurement/fine-tuning procedure iteratively allows the units to be tuned correctly even in the presence of slight imperfections further down the chain.

Simulations using the platform described above were used to validate the experimental flux signal propagation results.  The $f_z$ symmetry point of the target unit is defined as the unit's $f_z$ where the ground state expectation value of the z-loop current equals zero.  To track the magnitude of the flux signal, we recorded the difference in the target unit's z-symmetry point when the source unit was 20 m$\Phi_0$ on either side of the z-symmetry point.  This protocol was followed in both simulation and experiment.  Shown in Fig. 8, there is strong agreement between the two.

To measure the chain susceptibility, the entire data set was used, not just the source unit's $\pm$ 20 m$\Phi_0$ z-flux points.  The response of the target unit's $f_z$ symmetry point with respect to the source unit's $f_z$ was then fit to a sigmoid function.  The slope at the midpoint, $\frac{df^z_{c_1}}{df^z_{c_7}},$ was then extracted and used in the effective coupling strength calculation,
\begin{equation}
    \begin{split}
        J_{q_1 q_2}^{\rm{eff}}& = \chi_{c_1 c_7}(M_{q_1,c_1} I^z_{q_1})(M_{q_2,c_7} I^z_{q_2}) \\
        &\chi_{c_1 c_7} = \frac{d\langle I^z_{c_1} \rangle}{df_{c_7}^z} = \frac{d\langle I^z_{c_1} \rangle}{df_{c_1}^z} \frac{df^z_{c_1}}{df^z_{c_7}}.
    \end{split}
\end{equation}

To complete the cross chain effective strength calculation, single unit simulations were performed calculating the coupler's ground state expectation value of the z-loop current as well as the qubit's $\bra 0 I_z \ket 1$ matrix element at its $f_z$ symmetry point.  This measurement of $J_{\rm{eff}}$ was compared to a full device simulation of the qubit spectral line splitting and is displayed in Fig. 4c of the main text. 

\begin{figure*}[h]
\centering
\includegraphics[width = \linewidth]{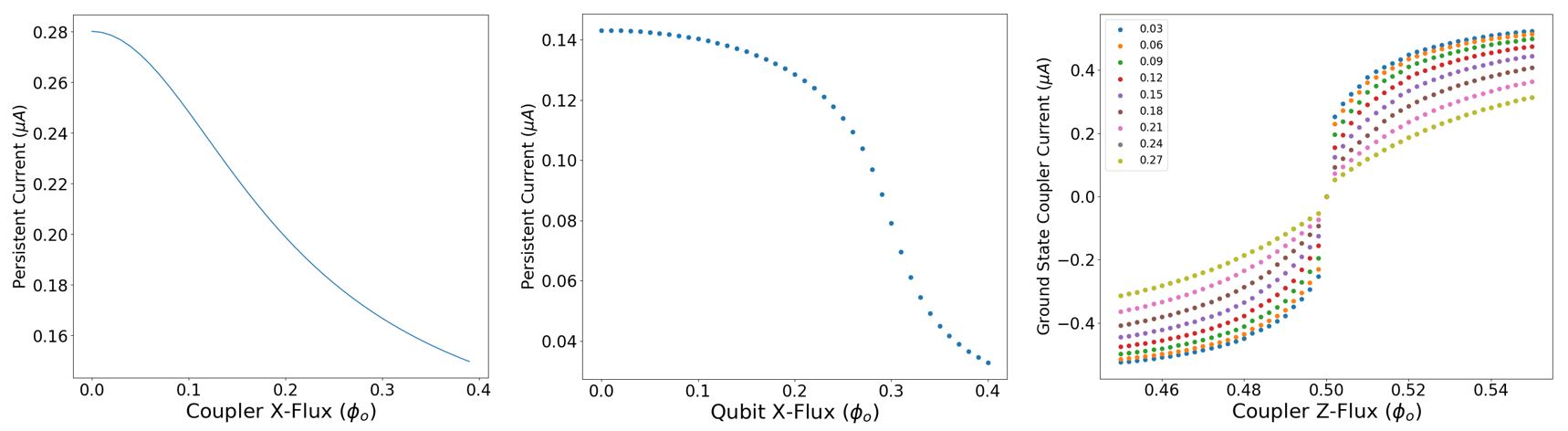}
\caption{\textbf{Calculation of $J^{\rm{eff}}_{q_1 q_2}$.} Left Panel.  The longitudinal coupling strength between coupler units, the persistent current of the coupler, $\bra 0 I_z \ket 1$, along the coupler's $f_z=\Phi_o/2$ was calculated by single coupler simulations.  Center Panel.  The persistent current of the qubit, $\bra 0 I_z \ket 1$, along the qubit's $f_z=\Phi_o/2$ was calculated by single qubit simulations. Right Panel.  The ground state expectation value of the z-loop current of the coupler, $\bra 0 I_z \ket 0$, was calculated for a number of coupler $f_x$ values, by single coupler simulations. This sweep was then fit to a sigmoid function and the slope at the z-loop symmetry point, $\frac{d\langle I^z_{c_1} \rangle}{df_{c_1}^z}$, was extracted.}

\end{figure*}

\newpage

\bibliography{SMbib}

